%% file: main.tex
\documentclass[11pt]{article}
\usepackage[preprint]{acl}
\usepackage{times,latexsym}
\usepackage[T1]{fontenc}
\usepackage{microtype}
\usepackage{amsmath,amssymb,graphicx,booktabs,array,tabularx,placeins}
\usepackage{natbib}
\usepackage{url}
\hypersetup{hidelinks,pdfauthor={Likai Wang, Dezhao Su, Jing Hu, Yunyun Wan, Jingdi Lei, Xinhai Li},pdftitle={From Tasks to Turns: Measuring Human Labor Contributions in Human-Agent Joint Production}}
\input{appendix_layout}
\title{From Tasks to Turns: Measuring Human Labor Contributions in Human--Agent Joint Production}
\input{authors}
\begin{document}
\maketitle
\authorrolenotes
\begin{abstract}
\input{abstract}
\end{abstract}
\input{chapters/01_introduction}
\input{chapters/02_related_work}
\input{chapters/figure_overview}
\input{chapters/03_method}
\input{chapters/04_experiments}
\input{chapters/05_results}
\input{chapters/06_discussion}
\input{chapters/07_conclusion}
\input{chapters/08_limitations_ethics}
\clearpage
\bibliography{bibliography/references}
\clearpage
\appendix
\startappendixflow
\clubpenalty=150
\widowpenalty=150
\input{appendix/appendix_experiment_tables}

\input{appendix/appendix_reliability}

\input{appendix/appendix_baselines}

\input{appendix/appendix_guidance}

\input{appendix/appendix_naturalness}

\end{document}

%% file: appendix_layout.tex
\usepackage{multicol}
\usepackage{xltabular}
\makeatletter
\newcommand{\startappendixflow}{%
  \onecolumn
  \raggedbottom
  \setlength{\multicolsep}{6pt plus 1pt minus 1pt}%
  \setlength{\premulticols}{40pt}%
  \setlength{\postmulticols}{12pt}%
  \setlength{\LTpre}{6pt}%
  \setlength{\LTpost}{8pt}%
  \setlength{\LTcapwidth}{\textwidth}%
}
\newenvironment{appendixtext}{%
  \begin{multicols}{2}%
}{%
  \par\end{multicols}%
}
\makeatother

%% file: authors.tex
\newcommand{\authorblock}[3]{%
  \begin{minipage}[t]{.30\textwidth}
    \centering\fontsize{9}{12}\selectfont
    {\fontsize{12}{12}\selectfont #1}\\
    #2\\
    {\fontsize{8}{12}\selectfont\texttt{#3}}\par
  \end{minipage}%
}
\author{%
  \normalfont
  \setlength{\tabcolsep}{0pt}%
  \begin{tabular}{@{}c@{\hspace{.015\textwidth}}c@{\hspace{.015\textwidth}}c@{}}
    \authorblock{Likai Wang\textsuperscript{*,$\dagger$}}{Human-in-the-Loop Co.,Ltd.\\Tsinghua University}{wlk26@mails.tsinghua.edu.cn} &
    \authorblock{Dezhao Su\textsuperscript{*}}{Fudan University}{dzsu25@m.fudan.edu.cn} &
    \authorblock{Jing Hu}{Peking University}{2400935232@stu.pku.edu.cn}\\\noalign{\vskip 10pt}
    \authorblock{Yunyun Wan}{Beijing University of\\Posts and Telecommunications}{wanyunyun@bupt.edu.cn} &
    \authorblock{Jingdi Lei}{Nanyang Technological University}{jingdi001@e.ntu.edu.sg} &
    \authorblock{Xinhai Li}{Harbin Institute of Technology}{22s151059@stu.hit.edu.cn}
  \end{tabular}%
}
\makeatletter
\newcommand{\authorrolenotes}{%
  \begingroup
  \renewcommand{\@makefntext}[1]{\noindent\@makefnmark##1}%
  \renewcommand{\thefootnote}{\fnsymbol{footnote}}%
  \footnotetext[1]{Equal contribution.\quad\textsuperscript{\textdagger}Corresponding author.}%
  \endgroup
}
\makeatother

%% file: abstract.tex
In agent workflows, a brief human instruction can alter subsequent search, reasoning, and tool execution, yet final task outcomes rarely reveal the contribution of an individual instruction to human--agent joint production. We propose a Human Turn-level framework for measuring human contributions, using task quality and token usage to characterize the collaborative production outcomes associated with each human instruction. We compare continuations from adjacent checkpoints under matched models, budgets, and termination conditions, and assess stability through repeated measurements. Across four archived and simulated interaction settings, we conduct 29,124 comparisons between predicted and observed directions of quality and computational change at both Human Turn and session levels. On HumanBench, we obtain 506 valid comparisons across 85 real-user scenarios. Results show that quality and computational changes associated with human guidance depend on task conditions. Expert-style and ordinary-style guidance differ in information provision, correction, and repetition. Even when task quality after continuation is identical, remaining computational demands can differ substantially across checkpoints. Our framework advances the measurement of human labor contributions from tasks and sessions to individual turns, providing a repeatable, fine-grained foundation for studying human contributions and their allocation in human--agent joint production.

%% file: chapters/01_introduction.tex
\section{Introduction}
\label{sec:introduction}

As language-model agents become involved in software development, tool use, and long-horizon tasks, human labor increasingly takes the form of natural-language guidance: identifying errors, adding constraints, supplying information, or deciding what should happen next. One brief instruction may save a lengthy search; another may prompt additional verification and tool calls. Its productive effect is reflected in the subsequent work jointly carried out by the human and agent. Measuring labor embedded in machine execution is therefore a fundamental problem for understanding human--agent joint production.

Task-based frameworks in labor economics explain technological change through the allocation of tasks between labor and capital \citep{acemoglu2019tasks}. Research on human--agent collaboration often evaluates final artifacts, task success, or entire sessions \citep{chang2025chatbench,shao2026collabskill}. These units capture team performance but aggregate human guidance, agent execution, and environmental feedback. Two sessions may pass the same acceptance test while consuming different amounts of computation, and individual guidance turns may accompany different forms of progress. Final outcomes alone cannot describe the production outcomes associated with individual human inputs.

We address this gap at the level of a single guidance turn, which we call a \emph{Human Turn}. We operationalize ``labor contributions'' not as hourly wages or monetary value, but as changes in joint execution state and remaining token use associated with human natural-language guidance. We record each human input alongside the subsequent joint work, retaining two separate dimensions: task quality, assessed against explicit acceptance criteria, and computational input, measured by logged token usage. Under matched task, model, and budget conditions, these dimensions connect guidance to observable outcomes and resources consumed during execution.

The framework combines session-level and turn-level records. For a session, we record final quality and the total tokens consumed by both executing and guiding models. For a Human Turn, we align the instruction and subsequent agent activity with adjacent checkpoints, independently resume the task from each checkpoint under matched conditions, and compare final quality and remaining tokens. These comparisons describe local collaborative state changes; they do not isolate the instruction's causal effect. We also examine guidance functions, information added beyond the preceding history, and task state to understand variation in the measurements.

We study five settings involving tool execution, code repair, mathematical reasoning, and human-origin conversations. Four archived and simulated settings yield 29,124 evaluation judgments comparing predicted and observed directions of quality and computational change. We compare expert-style and ordinary-style simulated guidance, assess perceived naturalness through blind human judgments, and examine stability through repeated autonomous continuations. HumanBench adds 506 valid checkpoint comparisons from 85 human-origin source sessions. Of these, 409 achieve identical final quality, while remaining token use varies in both directions.

Our contributions are threefold. First, we introduce a micro-level measurement framework organized around the Human Turn, linking human inputs to their associated joint execution. Second, we develop a two-dimensional protocol based on task acceptance and token usage, with matched conditions, complete model-token accounting, and repeated measurements. Third, we apply the framework to simulated guidance and human-origin interactions, exposing relationships between guidance behaviors, task states, and quality--computation outcomes that aggregate task results obscure. The framework supplies empirical quantities for studying labor contributions and collaboration capabilities, rather than assigning a monetary price to human labor.

%% file: chapters/02_related_work.tex
\section{Related Work}
\label{sec:related}

\subsection{Task Allocation and Human--Agent Evaluation}
Task-based economics studies allocation between labor and capital \citep{acemoglu2019tasks}; mixed-initiative and human--AI design work studies coordination and user control \citep{horvitz1999mixedinitiative,amershi2019guidelines}. Collaboration benchmarks emphasize joint outcomes or sessions \citep{chang2025chatbench,shao2026collaborativegym,shao2026collabskill}. Interaction logs and meta-analysis further reveal process detail and heterogeneous gains \citep{lee2022coauthor,vaccaro2024combinations}. We instead examine individual guidance turns and their associated execution.

\subsection{Language Feedback and Human Contributions}
Agents interleave reasoning, actions, and observations \citep{yao2023react}, and MINT studies simulated feedback in tool use \citep{wang2024mint}. Model critique and reflection support refinement \citep{madaan2023selfrefine,shinn2023reflexion,wadhwa2024refine}, yet multi-turn information can be used unreliably \citep{laban2026lost}. Unlike these feedback sources, our unit is a human turn. Whereas \citet{xie2026human} estimate human information in final artifacts, we pair guidance with subsequent quality and token use.

\subsection{Process Evaluation and Computational Input}
Executable benchmarks measure task completion \citep{liu2024agentbench,zhou2024webarena,jimenez2024swebench}; AgentBoard tracks execution progress \citep{ma2024agentboard}, and step verification supplies intermediate training supervision \citep{lightman2024verify}. Cost-aware evaluation and test-time scaling motivate measuring computation \citep{kapoor2024agents,snell2025compute}. We separately record full-session and checkpoint-to-finish tokens.

\subsection{Rubric-Based Evaluation and Simulated Users}
Rubrics and specialized evaluators make criteria explicit \citep{hashemi2024llmrubric,shao2026collabskill,liu2023geval,kim2024prometheus2}, but LLM judges have known biases \citep{zheng2023judge}. We therefore compare predictors with execution-derived evidence. Interaction benchmarks provide controlled simulated users \citep{zhou2024sotopia,yao2025taubench}, whose reliability requires separate study \citep{dou2025simulatorarena}. We assess linguistic naturalness blindly and execution effects on human-origin interactions.

%% file: chapters/figure_overview.tex
\begin{figure*}[t]
\centering
\includegraphics[width=\textwidth]{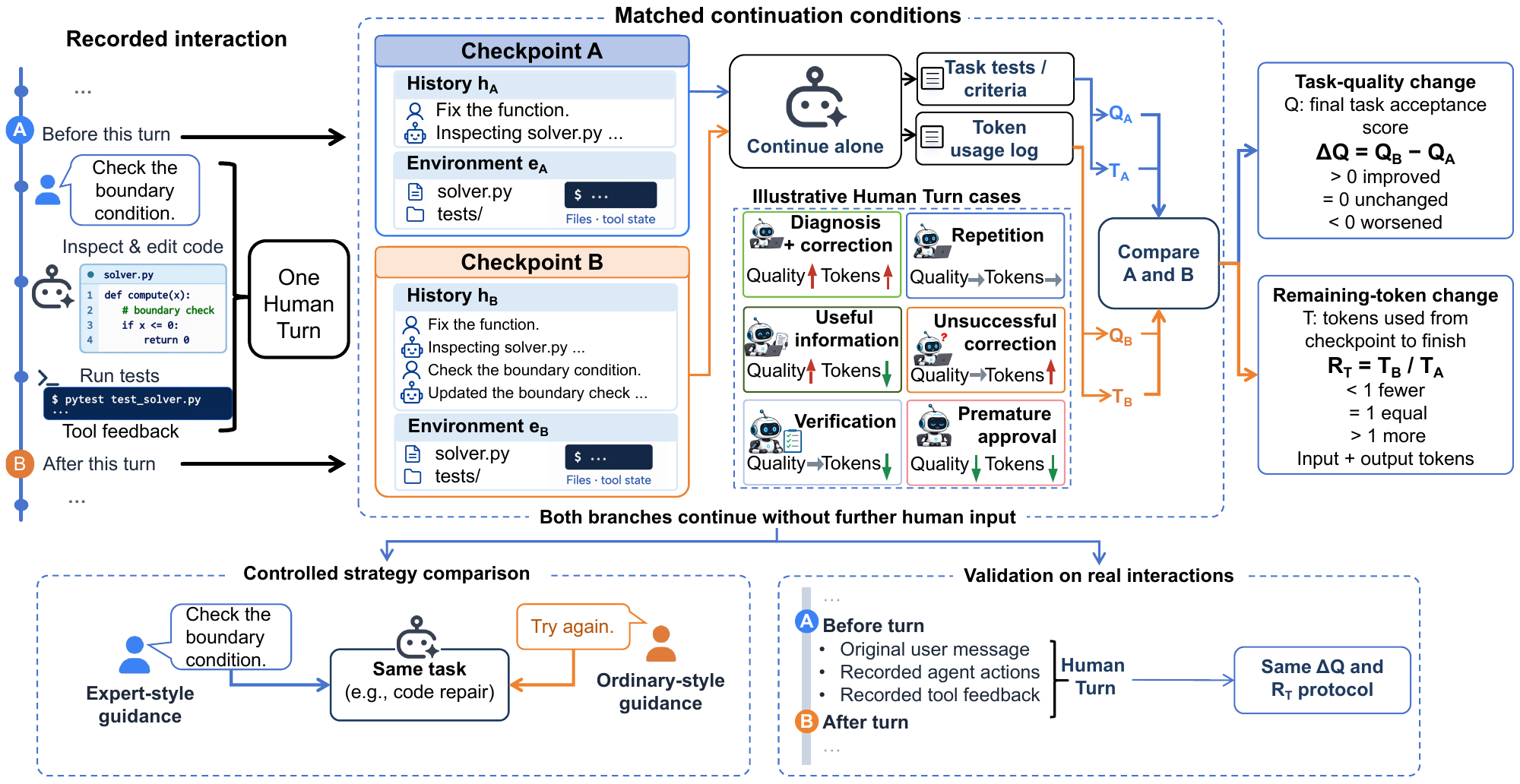}
\caption{Overview of the two-scale measurement framework. Under matched continuation conditions, adjacent checkpoints are independently continued to measure final task quality and remaining token usage. The framework also supports controlled guidance comparisons and validation on human-origin interactions.}
\label{fig:pipeline}
\end{figure*}

%% file: chapters/03_method.tex
\section{Problem Definition and Measurement}
\label{sec:measurement}

\subsection{Human Turn as the Unit of Analysis}
\label{sec:human-turn}
Let $g$ be the task objective and $u_t$ the human instruction at step $t$. Before the instruction, the state is $s_t=(g,h_t,e_t)$, where $h_t$ is the interaction history and $e_t$ the recoverable environment. A \emph{Human Turn} comprises the instruction and the joint work completed before the next interaction boundary:
\begin{equation}
H_t=(s_t,u_t,w_t,s_{t+1}),
\label{eq:human-turn}
\end{equation}
where $w_t$ includes agent execution and environmental feedback. Each unit is indexed by human input while retaining the subsequent state transition.

For archived interactions, we pair the recoverable checkpoint before an instruction ($A$) with its associated later checkpoint ($B$). The interval can include agent actions and tool feedback, so it measures joint state progression rather than an isolated instruction effect. Multiple instructions form a compound collaborative segment (Figure~\ref{fig:pipeline}).

\subsection{Task Quality and Remaining Computation}
\label{sec:quality-tokens}
At each endpoint $x\in\{A,B\}$, we restore the history and available environment, then independently continue the task under matched models, tools, budgets, and termination conditions. We record final task quality $Q_x$ against predefined task-specific criteria and remaining token usage $T_x$. Tokens include provider-reported input and output usage, without double-counting reasoning tokens already included in the total.
\begin{equation}
\Delta Q_t=Q_B-Q_A,\qquad R_{T,t}=\frac{T_B}{T_A},
\label{eq:quality-token-ratio}
\end{equation}
\begin{equation}
E_{T,t}=\log_2 R_{T,t}.
\label{eq:log-token-ratio}
\end{equation}
Positive, zero, and negative $\Delta Q_t$ indicate improved, unchanged, and decreased quality. Ties use equality of native scores with tolerance $10^{-12}$. For $T_A>0$, $E_{T,t}=-1$ indicates halved remaining tokens and $E_{T,t}=1$ indicates doubled remaining tokens.

Among quality-tied comparisons, $R_{T,t}<0.8$ denotes over 20\% fewer remaining tokens, $R_{T,t}>1.2$ denotes over 20\% more, and all other values form the middle category. These bands describe magnitude. A reduction in remaining computation is distinct from a net saving over the full task: the interval from $A$ to $B$ can itself consume tokens. We retain quality and computation separately because additional verification can increase both, whereas early termination can reduce both.

\subsection{Session Accounting and Collaboration Scorecards}
\label{sec:session-accounting}
At session level, we record final quality and total model tokens. With model-generated guidance,
\begin{equation}
C_{\mathrm{sys}}=C_a+C_u,
\label{eq:system-tokens}
\end{equation}
where $C_a$ and $C_u$ are execution-model and guidance-model tokens; independent execution has $C_u=0$. We aggregate valid repeated runs within each task, then weight tasks equally over the common valid set. Quality differences and relative changes in total tokens are reported separately.

The session ledger records consumed computation; $T_A$ and $T_B$ measure remaining work. Human time is not converted into model tokens. Scorecards retain these quantities separately; Appendices~\ref{app:coverage} and~\ref{app:repeats} give sample-flow and aggregation details.

\subsection{Comparable Conditions and Repeated Measurements}
\label{sec:comparability}
Valid comparisons require compatible task criteria, continuation models, and budgets. We record quality validity, token completeness, and technical interruptions separately. Joint analyses use their valid intersection; individual dimensions use their corresponding valid sets.

Each checkpoint receives one prespecified token estimate: the arithmetic mean of three complete, contract-matched continuations when available, or a previously selected valid single continuation otherwise. Comparisons sharing a checkpoint reuse that estimate. Repeated token measurements do not overwrite prespecified quality observations.

Repeatability analyses fix quality-tied comparisons with three complete continuations per endpoint. We compare individual runs and endpoint means, clustering bootstrap analyses by root task and checking alternative weighting schemes (Appendix~\ref{app:repeats}).

\subsection{Guidance Behavior and Text-Based Evaluation}
\label{sec:guidance-text}
We label simulated guidance by one primary linguistic function and potentially multiple information changes relative to its history (Appendix~\ref{app:guidance}).

We compare four text-based evaluators: Direct, without explicit criteria; CollabSkill-fixed, with eight fixed criteria \citep{shao2026collabskill}; RaR, with task-specific criteria but without reinforcement learning \citep{gunjal2026rar}; and JEV, with an independent model and native selection interface. Appendix~\ref{app:baselines} reports the complete per-setting evaluation.

Evaluators see the complete original session, but not independent continuation outcomes, measured continuation tokens, or reference labels. Their outputs are retrospective text-based judgments. Quality directions are decrease, tie, and increase; computational directions are decrease, middle, and increase, with $R_T\in[0.95,1.05]$ defining the middle class. This classification band differs from the 20\% magnitude bands above. We report accuracy, three-class Macro-F1, and coverage on common valid sets. Session-level judgments compare first and last checkpoints, rather than changes in the full session ledger.

%% file: chapters/04_experiments.tex
\section{Experimental Setup}
\label{sec:experimental-setup}

\subsection{Tasks and Datasets}
\label{sec:tasks}
We use five settings (Table~\ref{tab:settings}): multistep tool execution in Terminal-Bench \citep{merrill2026terminalbench}, code repair in HumanEvalFix \citep{muennighoff2024octopack}, two mathematical-reasoning settings based on Omni-MATH \citep{gao2025omnimath}, and human-origin code and text sessions in HumanBench. Omni100 and Omni300 overlap in underlying mathematics problems but follow different interaction and restoration protocols, so we report them separately.
\input{chapters/figure_turn_measurement}
\input{chapters/figure_guidance_semantics}
\input{chapters/figure_role_outcomes}
\input{chapters/table_settings}

Across four archived and simulated settings, we obtain 29,124 predicted-versus-observed directional judgments, summing four evaluators over quality and computation at adjacent- and first--last-checkpoint levels. Each dimension and level uses its common valid sample. Separately, 3,400 archived checkpoint pairs yield 3,168 token-complete, contract-matched pairs and 3,043 jointly valid quality--computation pairs. Session ledgers and evaluator outputs retain their own valid sets (Appendices~\ref{app:coverage} and~\ref{app:baselines}).

\subsection{Protocol, Baselines, and Reliability}
\label{sec:experimental-protocol}
We restore each checkpoint's history and available environment and continue to the specified termination conditions. Source-trajectory and continuation executors are registered separately; restoration batches and budget extensions retain their archived execution contracts. HumanEvalFix and Terminal source trajectories use GLM-4.7-Flash \citep{zai2026glm47flash,glm2025glm45}, with guidance from DeepSeek-V4-Flash \citep{deepseek2026v4}. Continuation assignments are detailed in Appendix~\ref{app:guidance}.

Text-based baselines test whether recorded language provides proxy estimates of production outcomes. Direct, CollabSkill-fixed, and RaR share the base prompt, output format, and evaluation model. Inapplicable criteria are ignorable soft evidence; RaR uses public task information. JEV has a separate model and interface, so its results also reflect evaluation-system differences.

Repeated measurements use a fixed subset of 1,075 quality-tied comparisons with three complete continuations per endpoint, covering 583 source sessions and 164 root tasks. We examine comparison-, session-, and task-weighted frequencies and use 10,000 root-task bootstrap resamples.

\subsection{Synthetic Collaboration Construction}
\label{sec:synthetic-guidance}
We generate expert-style and ordinary-style guidance with role-specific prompts in Terminal, HumanEvalFix, and Omni100. Assignments are verified against archived configurations and logs. Omni300 contains model teacher--student trajectories and is excluded from role comparisons.

For turn-level role comparisons, we match root tasks, source-generation conditions, and both continuation contracts. Token analysis additionally requires quality-tied comparisons for both roles within the same task and contract. We calculate proportions within task--contract--role groups, weight contracts equally within tasks, and weight root tasks equally. Intervals use 5,000 paired root-task bootstrap resamples. We retain archived linguistic-function and information-change labels with their separate valid denominators.

Ten evaluators experienced in agent programming assess the naturalness of original human utterances, prompt-based simulations, and utterances from a dialogue-adapted Qwen2.5-7B-Instruct model \citep{yang2024qwen25}. The adapted model uses WildChat and DevGPT exchanges \citep{zhao2024wildchat,xiao2024devgpt}. The blinded evaluation and overlap-stratified results appear in Appendix~\ref{app:naturalness}.

\subsection{Real Human Interaction Setup}
\label{sec:humanbench-setup}
HumanBench contains 85 human-origin source sessions covering code and general text tasks. User messages are linked to sessions, message indices, and checkpoints, yielding 600 pairs. Code tasks use task tests; text tasks use predefined binary requirement-satisfaction criteria, identical at both endpoints.

All 600 pairs have usable tokens; 94 lack valid quality observations, leaving 506 joint-valid pairs. These records come from 85 source sessions rather than 85 independent participants. Only aggregate execution-based results are reported; Appendix~\ref{app:humanbench-provenance} records the data boundary.

%% file: chapters/figure_turn_measurement.tex
\begin{figure*}[t]
\centering
\includegraphics[width=.93\textwidth,clip]{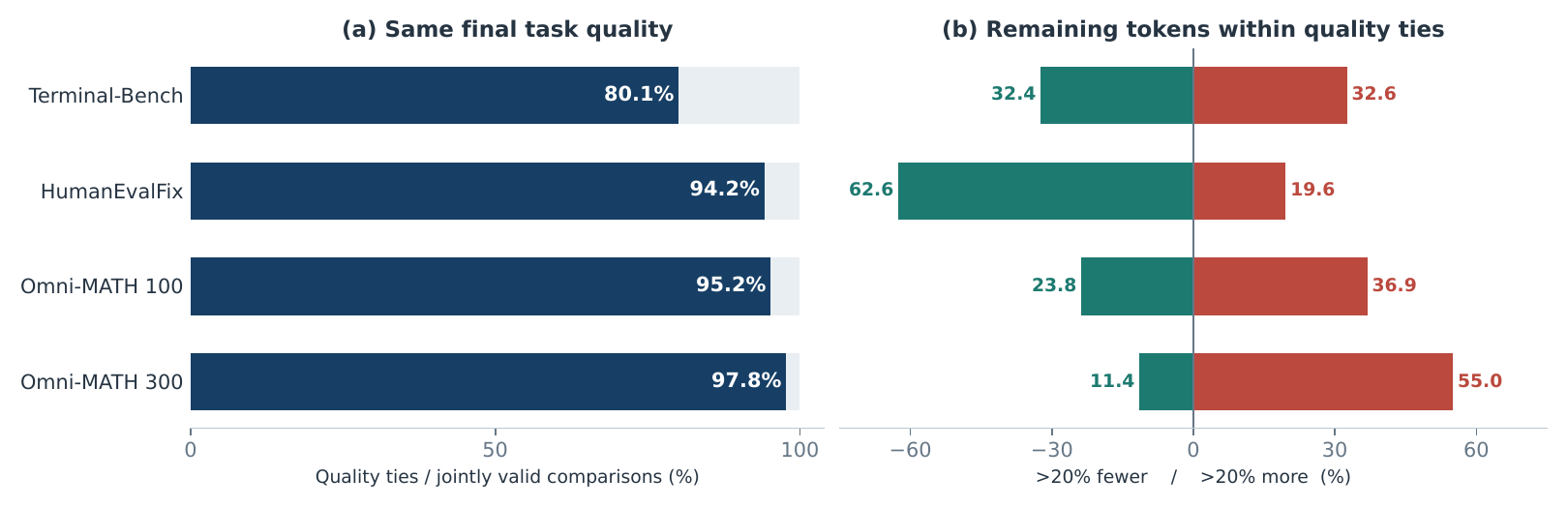}
\caption{Equal-quality frequencies and changes in remaining computation across four simulated or archived settings. Left-panel denominators are 644, 1,188, 168, and 537 jointly valid comparisons; right-panel denominators are the corresponding equal-quality subsets of 516, 1,119, 160, and 525. The right panel shows the proportions with $R_T<0.8$ and $R_T>1.2$; the middle category is omitted. All values use the prespecified unique cost estimate at each endpoint.}
\label{fig:turn_synthetic}
\end{figure*}

%% file: chapters/figure_guidance_semantics.tex
\begin{figure*}[t]
\centering
\includegraphics[width=.93\textwidth,clip]{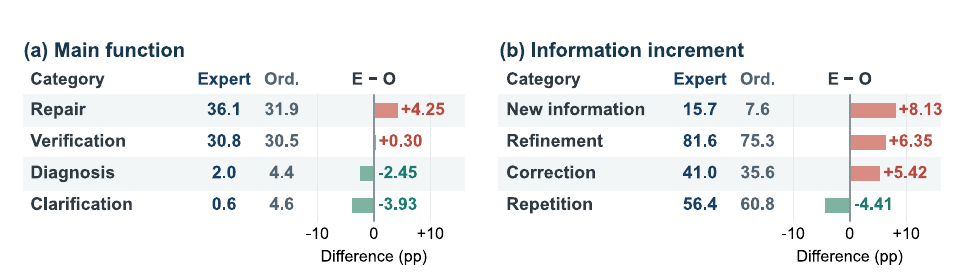}
\caption{Linguistic behaviors of expert-style and ordinary-style simulated guidance. Primary-function denominators are 1,866 and 2,055 turns, respectively; information-increment denominators are 1,131 and 1,261. The analysis covers HumanEvalFix, Terminal, and Omni100. Information increments may co-occur, so category percentages do not sum to 100\%. Bars show expert-style minus ordinary-style percentage-point differences.}
\label{fig:semantics}
\end{figure*}

%% file: chapters/figure_role_outcomes.tex
\begin{figure*}[t]
\centering
\includegraphics[width=.93\textwidth,clip]{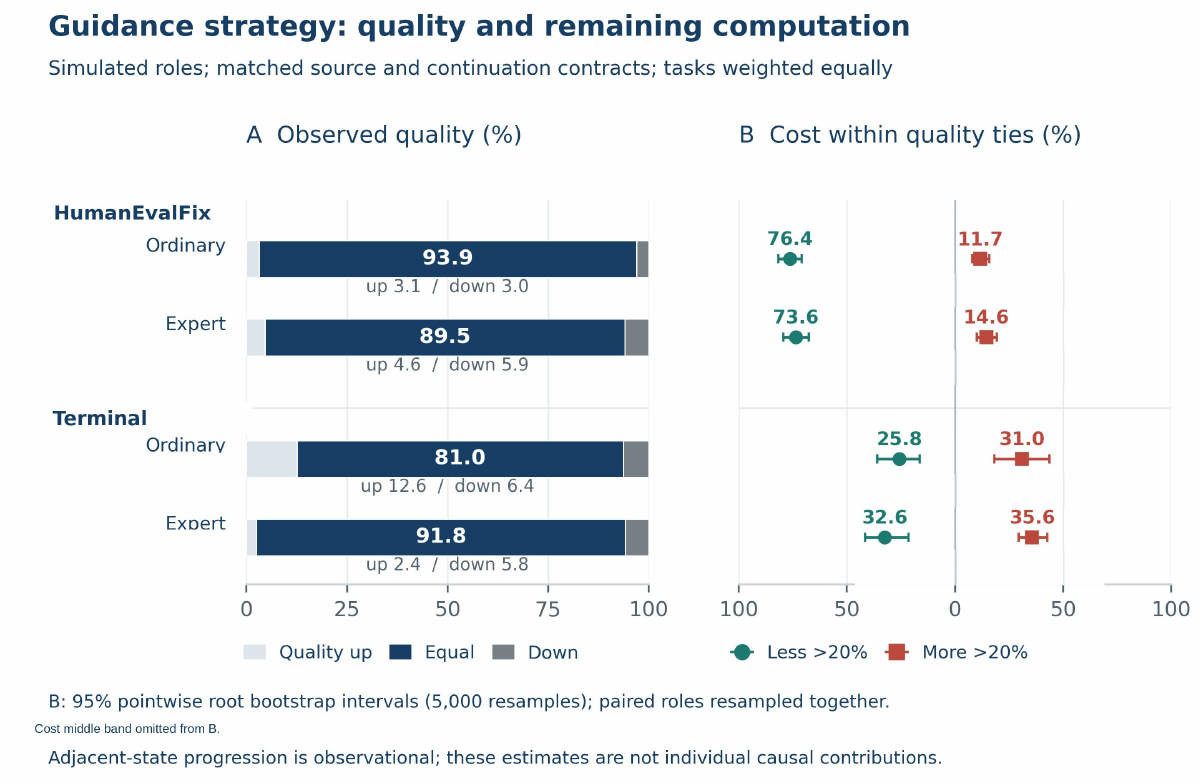}
\caption{Comparison of simulated guidance strategies under matched tasks, source configurations, and continuation conditions. The quality panel reports the directions of task-quality change, while the computation panel reports changes in remaining token consumption among quality-tied comparisons. Proportions are equally weighted across root tasks, with error bars representing pointwise 95\% confidence intervals from paired root-task bootstrap resampling. Turns are aggregated within matched task and execution conditions rather than paired by turn index.}
\label{fig:roleturn}
\end{figure*}

%% file: chapters/table_settings.tex
\begin{table*}[t]
\centering\small
\setlength{\tabcolsep}{4pt}
\begin{tabular}{@{}lp{3.6cm}p{3.4cm}rrr@{}}
\toprule
Setting & Guidance and task & Quality acceptance & Pairs & Joint-valid & Judgments \\
\midrule
Terminal & Simulated; tool execution & Native task acceptance & 734 & 644 & 5,072 \\
HumanEvalFix & Simulated; code repair & Binary test passing & 1,192 & 1,188 & 14,508 \\
Omni100 & Simulated; mathematics & Final-answer correctness & 192 & 168 & 2,096 \\
Omni300 & Archived teacher--student & Final-answer correctness & 682 & 537 & 7,448 \\
HumanBench & Human-origin code and text & Tests or predefined criteria & 600 & 506 & --- \\
\midrule
Total & & & 3,400 & 3,043 & 29,124 \\
\bottomrule
\end{tabular}
\caption{Five settings and evaluation scale. Pairs are adjacent checkpoint comparisons; joint validity requires valid quality, complete tokens, and matched execution contracts. Judgments sum four text evaluators over quality and computation at turn and first--last-checkpoint levels. HumanBench is outside that evaluator cohort. Appendix~\ref{app:coverage} details sample eligibility.}
\label{tab:settings}
\end{table*}

%% file: chapters/05_results.tex
\section{Results}
\label{sec:results}

\subsection{Why Human Turn-Level Measurement?}
\label{sec:whyturn}
Final quality alone does not capture computational demand. Across four archived and simulated settings, 2,537 checkpoint comparisons are jointly valid and 2,320 achieve identical final quality. Tie rates range from 80.1\% to 97.8\%, yet remaining tokens vary in both directions within these ties (Figure~\ref{fig:turn_synthetic}). In HumanEvalFix, 62.6\% of tied comparisons use over 20\% fewer remaining tokens and 19.6\% use over 20\% more. The corresponding proportions are 11.4\% and 55.0\% in Omni300, and 32.4\% and 32.6\% in Terminal. Quality and computational input therefore provide complementary observations of collaborative production.

Text-based judgments provide proxy estimates. Table~\ref{tab:baselines} compares predictions of remaining-computation direction. At turn level, JEV has the highest average accuracy (53.3\%), while RaR has the highest three-class Macro-F1 (36.1\%). At session level, CollabSkill-fixed has the highest accuracy (55.7\%) and RaR the highest Macro-F1 (35.1\%). Different rankings distinguish overall accuracy from balanced recognition of all three directions.
\input{chapters/table_baselines}

Repeated autonomous continuations under matched archived configurations exhibit consistent aggregate patterns (Table~\ref{tab:repeats}). On the fixed quality-tied subset, three runs yield proportions of 60.93\%, 61.67\%, and 61.02\% with over 20\% fewer tokens, and 20.19\%, 20.84\%, and 22.60\% with over 20\% more. Averaging tokens at each endpoint before computing ratios gives 62.51\% and 19.26\%, respectively.
\input{chapters/table_repeats}

\subsection{Comparing Synthetic Guidance Strategies}
\label{sec:strategies}
The strategies differ in linguistic behavior (Figure~\ref{fig:semantics}). Expert-style guidance more often supplies new information (15.7\% versus 7.6\%), refinements (81.6\% versus 75.3\%), and corrections (41.0\% versus 35.6\%). Ordinary-style guidance more often contains repetition (60.8\% versus 56.4\%) and clarification (4.6\% versus 0.6\%). Role-specific prompting thus produces different information-provision and interaction patterns.

Guidance-associated outcomes depend on task conditions. Figure~\ref{fig:roleturn} compares matched tasks and execution conditions. Among quality-tied HumanEvalFix comparisons, the task-weighted proportions with over 20\% fewer remaining tokens are 76.37\% for ordinary-style and 73.56\% for expert-style guidance. Terminal yields 25.79\% and 32.61\%, respectively.

Complete-session outcomes provide a complementary view (Table~\ref{tab:role_session}). In Terminal, expert-style guidance achieves 81.04\% quality versus 70.65\% for ordinary-style guidance, with 4.0\% fewer total tokens in the point estimates. HumanEvalFix yields 98.57\% versus 100.00\% quality, with 7.8\% more expert-style tokens. In Omni100, both achieve 100.00\% quality, while expert-style guidance uses 10.6\% fewer tokens. These descriptive estimates retain the archived paired-session cohorts.
\input{chapters/table_role_session}

Complete accounting distinguishes execution from guidance tokens. In HumanEvalFix, independent execution averages 16.77k tokens. Collaboration uses 19.53k execution tokens plus 14.63k guidance tokens: 34.16k in total, a 103.7\% increase. Guidance contributes 42.8\% of collaborative model tokens. The session ledger therefore captures resources that local checkpoint measurements do not represent.

In the blinded naturalness study's subset without exact-match exchanges, original human messages, prompt-based simulations, and dialogue-adapted utterances are selected as most human-like in 32.1\%, 26.7\%, and 41.2\% of valid judgments. This provides a human assessment of linguistic naturalness (Appendix~\ref{app:naturalness}).

\subsection{Validation on Real Human Interactions}
\label{sec:humanresults}
HumanBench provides 506 valid comparisons across 85 human-origin source sessions. Of these, 49 (9.7\%) improve quality, 409 (80.8\%) preserve quality, and 48 (9.5\%) decrease quality (Figure~\ref{fig:human}).
\input{chapters/figure_humanbench}

Identical quality can conceal different remaining work. Among the 409 tied comparisons, 112 (27.4\%) use over 20\% fewer remaining tokens, 128 (31.3\%) use over 20\% more, and 169 (41.3\%) fall in the middle band. Thus, 58.7\% exhibit token changes exceeding 20\% that final task quality alone does not distinguish. These observations extend the two-dimensional protocol to human-origin guidance and provide a fine-grained basis for examining labor contributions in joint production.

%% file: chapters/table_baselines.tex
\begin{table}[t]
\centering
\small
\setlength{\tabcolsep}{3pt}
\begin{tabular}{@{}lrrrr@{}}
\toprule
& \multicolumn{2}{c}{Turn} & \multicolumn{2}{c}{First--last} \\
\cmidrule(lr){2-3}\cmidrule(l){4-5}
Method & Acc & F1 & Acc & F1 \\
\midrule
Direct & 43.3 & 34.8 & 45.3 & 32.5 \\
CollabSkill-fixed & 48.2 & 34.9 & 55.7 & 34.8 \\
RaR & 47.5 & 36.1 & 51.2 & 35.1 \\
JEV & 53.3 & 34.1 & 54.8 & 28.1 \\
\bottomrule
\end{tabular}
\caption{Text-based judgments of remaining-computation direction, averaged with equal weight across four settings (\%). F1 denotes three-class Macro-F1. First--last columns compare checkpoint continuations. The middle cost category is $R_T\in[0.95,1.05]$. Results retain the original evaluation sample; per-setting results and per-axis denominators appear in Appendix~\ref{app:baselines}.}
\label{tab:baselines}
\end{table}

%% file: chapters/table_repeats.tex
\begin{table}[t]
\centering
\small
\setlength{\tabcolsep}{4pt}
\renewcommand{\arraystretch}{1.1}
\begin{tabular}{@{}lrr@{}}
\toprule
Token estimate & \shortstack[r]{Savings\\$>20\%$} & \shortstack[r]{Increase\\$>20\%$} \\
\midrule
Continuation 1 & 60.93 & 20.19 \\
Continuation 2 & 61.67 & 20.84 \\
Continuation 3 & 61.02 & 22.60 \\
Three-run endpoint mean & 62.51 & 19.26 \\
\bottomrule
\end{tabular}
\caption{Repeated-measurement frequencies (\%) on a fixed subset of 1,075 equal-quality comparisons. The mean row first averages token usage across three continuations at each endpoint, then computes the ratio and category.}
\label{tab:repeats}
\end{table}

%% file: chapters/table_role_session.tex
\begin{table*}[t]
\centering
\small
\setlength{\tabcolsep}{5pt}
\renewcommand{\arraystretch}{1.15}
\begin{tabular}{@{}lrrrr@{}}
\toprule
Setting & \shortstack[r]{Ordinary\\quality (\%)} & \shortstack[r]{Expert\\quality (\%)} & \shortstack[r]{Quality difference\\(pp)} & \shortstack[r]{Total-token change\\(\%)} \\
\midrule
HumanEvalFix & 100.00 & 98.57 & $-1.43$ & $+7.8$ \\
Omni100 & 100.00 & 100.00 & $0.00$ & $-10.6$ \\
Terminal & 70.65 & 81.04 & $+10.39$ & $-4.0$ \\
\bottomrule
\end{tabular}
\caption{Full-session point estimates for simulated roles on the original paired cohorts. Quality differences are expert-style minus ordinary-style; total-token changes are expert-style relative to ordinary-style.}
\label{tab:role_session}
\end{table*}

%% file: chapters/figure_humanbench.tex
\begin{figure*}[t]
\centering
\includegraphics[width=.93\textwidth,clip]{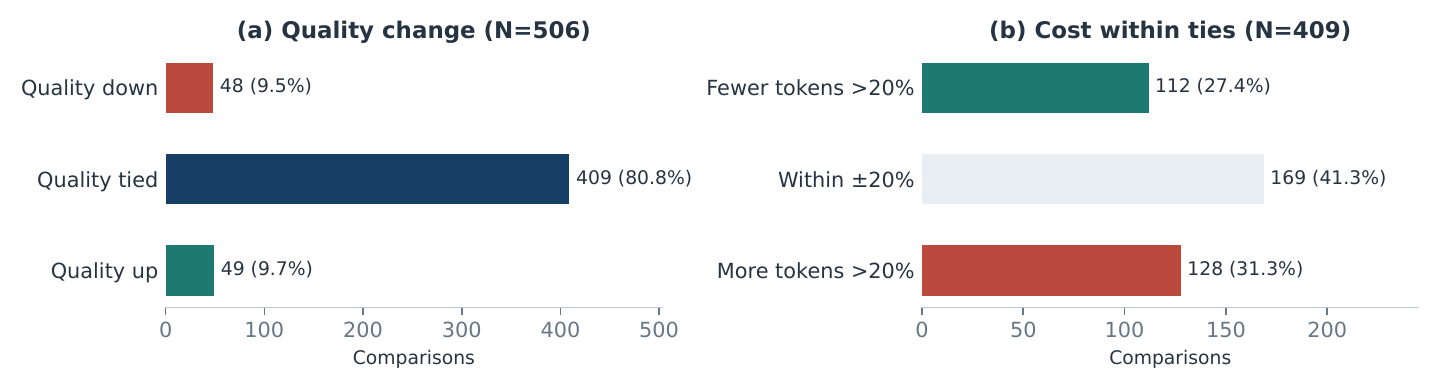}
\caption{Execution-based measurements on human-origin HumanBench interactions. Panel (a) shows quality directions for 506 jointly valid comparisons. Panel (b) shows remaining computation for the 409 equal-quality comparisons, including 112 with more than 20\% fewer tokens, 169 within $\pm20\%$, and 128 with more than 20\% additional tokens. The equal-quality subset covers 48 root tasks and 82 source sessions.}
\label{fig:human}
\end{figure*}

%% file: chapters/06_discussion.tex
\section{Discussion}
\label{sec:discussion}

\subsection{From Task Outcomes to Human Labor Contributions}
Human guidance supplies information, corrects errors, and redirects machine execution. Human Turn-level records connect each instruction to the associated joint work, acceptance outcomes, and remaining computation. Alongside session accounting, these records expose local production differences without collapsing collaboration into an overall score.

\subsection{Comparable Measurement of Quality and Computation}
Task acceptance and logged tokens make quality and computation observable. Matched execution conditions, complete accounting, and repeated continuations establish comparability without imposing an exchange rate between the two dimensions. Here, fairness means operational comparability, rather than a definitive valuation of labor or attribution of net contribution to an instruction.

\subsection{Studying Value in Human--Agent Joint Production}
These records supply empirical variables for studying guidance skills, division of labor, and resource allocation. Comparing people requires broader shared-task coverage, repeated observations, and controls for effort. Controlled counterfactual continuations, human time, and outcome value could support future studies of attributable contributions and value distribution.

%% file: chapters/07_conclusion.tex
\section{Conclusion}
\label{sec:conclusion}
A human instruction can shape subsequent search, reasoning, tool use, and task execution. We introduce a Human Turn-level framework that associates this joint work with task quality and computational input. Across five settings, complete session accounting, adjacent-checkpoint continuations, and guidance analysis expose process-level differences obscured by final outcomes. In particular, 506 valid comparisons from 85 human-origin source sessions show that remaining computation can differ despite identical task acceptance. Moving from tasks and sessions to Human Turns establishes a reproducible micro-level empirical basis for studying the productive roles of guidance, comparing collaborative behaviors, and investigating labor contributions in human--agent joint production.

%% file: chapters/08_limitations_ethics.tex
\section*{Limitations}
\label{sec:limitations}

\paragraph{Measurement and attribution.}
Our framework measures collaborative state changes associated with human guidance. Adjacent-checkpoint intervals may contain human instructions, agent execution, and environmental feedback, so the observed changes cannot be interpreted as isolated causal effects of individual instructions. Equal native acceptance scores do not necessarily imply identical underlying quality. Remaining computation differs from total session resources, and token usage cannot be equated with the economic value of human labor. The framework provides an empirical measurement basis rather than a complete causal or economic valuation model.

\paragraph{Data and scope.}
Expert-style and ordinary-style guidance are simulated strategies. Blind human evaluation does not make their differences directly generalizable to real human populations. HumanBench contains genuine instructions from 85 source sessions, not 85 independent participants. Our settings cover software engineering, multistep tool execution, mathematical reasoning, and general interactions. Applicability to other domains requires further validation.

\paragraph{Comparability and generalization.}
Task criteria, models, tokenizers, budgets, and termination conditions define the scope of comparison. Repeated measurements assess stability, but stochastic execution can still affect individual pairs. Observations from different tasks and conditions are not inherently comparable. The protocol applies primarily to joint production with explicit objectives, verifiable criteria, and observable computation. Comparing real individuals requires broader shared-task coverage, repeated observations, and controls for human effort and collaboration conditions. Settings without standardized acceptance criteria, including creative writing, open-ended decisions, and social interaction, require further study.

\section*{Ethical Considerations}
\label{sec:ethics}
We report anonymized aggregate statistics for human-origin interactions. Original private messages, proprietary code, and identifying information are not distributed with the paper. Any later release of interactions requires separate verification of source permissions and redistribution rights.

This work develops an empirical basis for studying human labor contributions, rather than directly determining wages, hiring decisions, or performance evaluations. Comparisons between people must account for tasks, computational resources, experience, and uncertainty. Local computational changes should not be treated as complete assessments of individual labor value.

%% file: appendix/appendix_experiment_tables.tex
\begin{appendixtext}
\section{Sample Eligibility and Measurement Units}\label{app:coverage}

Table~\ref{tab:strict_all} reports the successive denominators for the strict joint analysis of quality and computation. Equal-quality proportions use jointly valid comparisons as their denominator; savings and increase proportions use equal-quality comparisons.
\end{appendixtext}

\begingroup\small
\begin{xltabular}{\textwidth}{@{}>{\raggedright\arraybackslash}Xrrrrr@{}}
\caption{Strict sample flow and computation categories at equal quality. Savings / middle / increase correspond to $R_T<0.8$, $0.8\le R_T\le1.2$, and $R_T>1.2$, respectively. All entries are counts of comparisons.}\label{tab:strict_all}\\
\toprule
\textbf{Setting} & \textbf{Raw} & \textbf{\shortstack{Cost-\\eligible}} & \textbf{\shortstack{Jointly\\valid}} & \textbf{\shortstack{Equal\\quality}} & \textbf{\shortstack{Savings / middle\\/ increase}} \\
\midrule
\endfirsthead
\multicolumn{6}{@{}l}{Table~\thetable\ (continued)}\\
\toprule
\textbf{Setting} & \textbf{Raw} & \textbf{\shortstack{Cost-\\eligible}} & \textbf{\shortstack{Jointly\\valid}} & \textbf{\shortstack{Equal\\quality}} & \textbf{\shortstack{Savings / middle\\/ increase}} \\
\midrule
\endhead
\midrule\multicolumn{6}{r@{}}{\textit{Continued on next page}}\\
\endfoot
\bottomrule
\endlastfoot
Terminal & 734 & 647 & 644 & 516 & 167 / 181 / 168 \\
HumanEvalFix & 1,192 & 1,188 & 1,188 & 1,119 & 700 / 200 / 219 \\
Omni100 & 192 & 168 & 168 & 160 & 38 / 63 / 59 \\
HumanBench & 600 & 600 & 506 & 409 & 112 / 169 / 128 \\
Omni300 & 682 & 565 & 537 & 525 & 60 / 176 / 289 \\
Total & 3,400 & 3,168 & 3,043 & 2,729 & 1,077 / 789 / 863 \\
\end{xltabular}
\endgroup

\begin{appendixtext}
The original quality inventory (Table~\ref{tab:quality_inventory}) retains the historical quality-valid sample and addresses a different eligibility criterion. Omni100 and Omni300 have 60 and 297 original root tasks, respectively, whereas strict joint validity covers 59 and 264; these counts are not interchangeable.

Quality retains the originally designated observation, while cost uses one estimate per node. Whether a cost record uses a repeat mean or a single observation depends on complete repeat groups and contract eligibility. Invalid quality cannot be assigned to failure or no change by default; records with complete cost but invalid quality are eligible only for cost-only analyses. Restored original budgets are interpreted under the contract actually bound to the run, and different restoration conditions are not pooled according to their outcomes.
\end{appendixtext}

\begingroup\small
\begin{xltabular}{\textwidth}{@{}>{\raggedright\arraybackslash}Xrrrr@{}}
\caption{Original quality inventory. Quality-valid and equal-quality counts belong to the original quality-observation sample and do not replace the strict joint quality-and-cost denominators in Table~\ref{tab:strict_all}.}\label{tab:quality_inventory}\\
\toprule
\textbf{Setting} & \textbf{\shortstack{Original\\root tasks}} & \textbf{\shortstack{Adjacent\\comparisons}} & \textbf{\shortstack{Quality-\\valid}} & \textbf{\shortstack{Quality decrease /\\unchanged / increase}} \\
\midrule
\endfirsthead
\multicolumn{5}{@{}l}{Table~\thetable\ (continued)}\\
\toprule
\textbf{Setting} & \textbf{\shortstack{Original\\root tasks}} & \textbf{\shortstack{Adjacent\\comparisons}} & \textbf{\shortstack{Quality-\\valid}} & \textbf{\shortstack{Quality decrease /\\unchanged / increase}} \\
\midrule
\endhead
\midrule\multicolumn{5}{r@{}}{\textit{Continued on next page}}\\
\endfoot
\bottomrule
\endlastfoot
Terminal & 22 & 734 & 731 & 73/573/85 \\
HumanEvalFix & 164 & 1,192 & 1,188 & 37/1,119/32 \\
Omni100 & 60 & 192 & 192 & 8/180/4 \\
HumanBench & 49 & 600 & 506 & 48/409/49 \\
Omni300 & 297 & 682 & 629 & 6/611/12 \\
\end{xltabular}
\endgroup

%% file: appendix/appendix_reliability.tex
\begin{figure}[!htb]
\centering
\includegraphics[width=.88\textwidth]{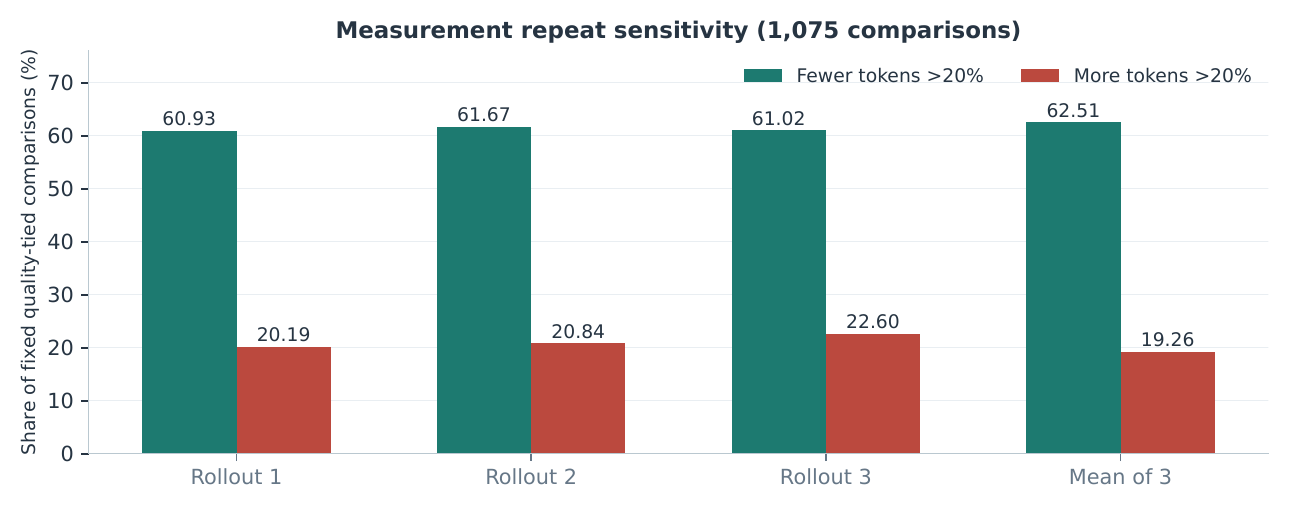}
\caption{Repeated measurements on a fixed subset of 1,075 quality-tied comparisons. Each endpoint has three complete continuations. Individual runs and the three-run endpoint mean show the proportions with more than 20\% fewer or more remaining tokens. Endpoint means are computed before forming token ratios.}
\label{fig:repeat}
\end{figure}

\begin{appendixtext}
\section{Repeatability and Weighting Sensitivity}\label{app:repeats}

We measure repeatability on a fixed subset of equal-quality comparisons. Eligibility requires equal original quality, complete costs at both endpoints, and three continuations matching the contract at each endpoint. Of 1,119 original equal-quality comparisons, 1,075 meet these requirements; the remaining 44 lack complete triplicates at both endpoints. Each repeat and the endpoint-mean calculation use this same eligible subset.

With equal weight per source session on this subset, savings frequencies range from 65.80\%--68.51\% across the three repeats, and increase frequencies range from 15.40\%--19.14\%. With equal weight per root task, the corresponding ranges are 63.83\%--64.63\% and 17.77\%--21.20\%. Leaving out one continuation at a time and averaging the remaining two at each endpoint gives savings rates of 60.00\%, 60.93\%, and 62.79\%, and increase rates of 20.19\%, 19.63\%, and 18.88\%. In all three cases, bootstrap intervals over root tasks for the difference between savings and increase frequencies remain positive.

For variation in single-run tokens at the same checkpoint, 1,797 complete states yield 5,391 pairwise ratios, with an empirical 90\% ratio-coverage scale of approximately 2.977. Thus, 20\% serves as a display threshold for magnitude, not a test of statistical significance for an individual comparison.

Figure~\ref{fig:repeat} summarizes the comparison-weighted results. With equal weight per comparison, the three continuations yield savings frequencies of 60.93\%, 61.67\%, and 61.02\%, and increase frequencies of 20.19\%, 20.84\%, and 22.60\%, respectively. Averaging all three token measurements at each endpoint before computing ratios yields 62.51\% savings and 19.26\% increases. All estimates use the same 1,075 comparisons.
\end{appendixtext}

%% file: appendix/appendix_baselines.tex
\begin{appendixtext}
\section{Results of Text-Based Evaluation by Setting}\label{app:baselines}

All four methods receive the complete original conversation, including its later original turns. New continuations, measured tokens, reference labels, and hidden acceptance information are withheld from the evaluators. Direct receives no additional criteria; CollabSkill-fixed uses eight fixed criteria; RaR generates and reuses task criteria from the public task and continuation contract; JEV uses its native choice interface and fixed criteria. Execution outcomes beyond public task information are excluded from criterion generation.

Quality and cost use separate validity masks, and coverage statistics include technical failures. For text-based cost judgments, $R_T\in[0.95,1.05]$ defines the middle class; this classification band is distinct from the main analysis's 20\% magnitude bands. Three-class macro-F1 always includes all labels to make majority-class effects visible alongside accuracy.

\subsection{Adjacent-Checkpoint Judgments}
Table~\ref{tab:turn_baselines} reports turn-level results on common support in each setting.
\end{appendixtext}

\begingroup\small
\begin{xltabular}{\textwidth}{@{}>{\raggedright\arraybackslash}Xlrrr@{}}
\caption{Turn-level direction judgments from four text-based methods by setting. $N_Q$ and $N_T$ are the common-support counts for quality and cost. Accuracy and three-class macro-F1 use the frozen evaluation sample, which differs from the strictly jointly valid sample in the main analysis.}\label{tab:turn_baselines}\\
\toprule
\textbf{Setting} & \textbf{Method} & \textbf{$N_Q/N_T$} & \textbf{\shortstack{Quality\\Acc/Macro-F1 (\%)}} & \textbf{\shortstack{Cost\\Acc/Macro-F1 (\%)}} \\
\midrule
\endfirsthead
\multicolumn{5}{@{}l}{Table~\thetable\ (continued)}\\
\toprule
\textbf{Setting} & \textbf{Method} & \textbf{$N_Q/N_T$} & \textbf{\shortstack{Quality\\Acc/Macro-F1 (\%)}} & \textbf{\shortstack{Cost\\Acc/Macro-F1 (\%)}} \\
\midrule
\endhead
\midrule\multicolumn{5}{r@{}}{\textit{Continued on next page}}\\
\endfoot
\bottomrule
\endlastfoot
Terminal & Direct & 546/513 & 60.8/32.2 & 35.5/30.8 \\
Terminal & CollabSkill-fixed & 546/513 & 55.7/31.5 & 36.5/29.5 \\
Terminal & RaR & 546/513 & 63.7/34.0 & 41.7/32.5 \\
Terminal & JEV & 546/513 & 27.5/18.4 & 45.4/30.5 \\
\addlinespace
HumanEvalFix & Direct & 1254/1192 & 87.6/32.9 & 52.9/39.1 \\
HumanEvalFix & CollabSkill-fixed & 1254/1192 & 88.2/32.5 & 57.0/41.1 \\
HumanEvalFix & RaR & 1254/1192 & 86.3/32.9 & 53.7/38.9 \\
HumanEvalFix & JEV & 1254/1192 & 80.5/35.2 & 51.1/35.9 \\
\addlinespace
Omni100 & Direct & 185/185 & 85.4/30.8 & 35.7/33.6 \\
Omni100 & CollabSkill-fixed & 185/185 & 87.6/31.2 & 39.5/36.4 \\
Omni100 & RaR & 185/185 & 87.0/31.1 & 41.6/38.0 \\
Omni100 & JEV & 185/185 & 81.6/32.8 & 50.8/39.1 \\
\addlinespace
Omni300 & Direct & 625/678 & 84.6/30.6 & 49.0/35.9 \\
Omni300 & CollabSkill-fixed & 625/678 & 84.6/30.6 & 60.0/32.7 \\
Omni300 & RaR & 625/678 & 85.4/30.7 & 52.9/35.1 \\
Omni300 & JEV & 625/678 & 87.4/31.1 & 65.8/30.9 \\
\end{xltabular}
\endgroup

\begin{appendixtext}
\subsection{First--Last-Checkpoint Judgments}
Table~\ref{tab:session_baselines} reports session-level direction judgments on common support in each setting. These judgments compare checkpoint continuations, rather than total session bills.
\end{appendixtext}

\begingroup\small
\begin{xltabular}{\textwidth}{@{}>{\raggedright\arraybackslash}Xlrrr@{}}
\caption{Direction judgments between the first and last checkpoints from four text-based methods, by setting. Denominators and scores retain the frozen sample. Session-level direction does not represent a change in the total session bill.}\label{tab:session_baselines}\\
\toprule
\textbf{Setting} & \textbf{Method} & \textbf{$N_Q/N_T$} & \textbf{\shortstack{Quality\\Acc/Macro-F1 (\%)}} & \textbf{\shortstack{Cost\\Acc/Macro-F1 (\%)}} \\
\midrule
\endfirsthead
\multicolumn{5}{@{}l}{Table~\thetable\ (continued)}\\
\toprule
\textbf{Setting} & \textbf{Method} & \textbf{$N_Q/N_T$} & \textbf{\shortstack{Quality\\Acc/Macro-F1 (\%)}} & \textbf{\shortstack{Cost\\Acc/Macro-F1 (\%)}} \\
\midrule
\endhead
\midrule\multicolumn{5}{r@{}}{\textit{Continued on next page}}\\
\endfoot
\bottomrule
\endlastfoot
Terminal & Direct & 108/101 & 43.5/31.5 & 42.6/36.7 \\
Terminal & CollabSkill-fixed & 108/101 & 41.7/30.7 & 41.6/32.1 \\
Terminal & RaR & 108/101 & 39.8/29.1 & 43.6/31.0 \\
Terminal & JEV & 108/101 & 21.3/19.3 & 39.6/18.9 \\
\addlinespace
HumanEvalFix & Direct & 601/580 & 84.2/33.6 & 44.1/31.6 \\
HumanEvalFix & CollabSkill-fixed & 601/580 & 84.9/31.6 & 48.3/32.6 \\
HumanEvalFix & RaR & 601/580 & 81.0/32.4 & 46.2/30.4 \\
HumanEvalFix & JEV & 601/580 & 68.6/32.2 & 31.7/20.9 \\
\addlinespace
Omni100 & Direct & 77/77 & 80.5/30.0 & 39.0/29.7 \\
Omni100 & CollabSkill-fixed & 77/77 & 83.1/30.5 & 51.9/42.8 \\
Omni100 & RaR & 77/77 & 81.8/30.2 & 49.4/40.8 \\
Omni100 & JEV & 77/77 & 76.6/28.9 & 66.2/42.6 \\
\addlinespace
Omni300 & Direct & 272/287 & 68.0/27.0 & 55.4/32.0 \\
Omni300 & CollabSkill-fixed & 272/287 & 68.0/27.0 & 81.2/31.5 \\
Omni300 & RaR & 272/287 & 69.9/27.4 & 65.9/38.1 \\
Omni300 & JEV & 272/287 & 73.9/28.3 & 81.5/29.9 \\
\end{xltabular}
\endgroup

\begin{appendixtext}
\subsection{Label Distributions}
Table~\ref{tab:majority} reports the quality and cost label distributions. When binary quality scores approach their ceiling, the unchanged class can dominate, so accuracy should be interpreted together with these distributions.
\end{appendixtext}

\begingroup\small
\begin{xltabular}{\textwidth}{@{}>{\raggedright\arraybackslash}Xlrrrr@{}}
\caption{Majority classes and label distributions in the original evaluation sample, for interpreting accuracy and macro-F1. Quality and cost each use their own valid sample.}\label{tab:majority}\\
\toprule
\textbf{Setting} & \textbf{Dimension} & \textbf{Decrease} & \textbf{\shortstack{Unchanged\\/ middle}} & \textbf{Increase} & \textbf{$N$} \\
\midrule
\endfirsthead
\multicolumn{6}{@{}l}{Table~\thetable\ (continued)}\\
\toprule
\textbf{Setting} & \textbf{Dimension} & \textbf{Decrease} & \textbf{\shortstack{Unchanged\\/ middle}} & \textbf{Increase} & \textbf{$N$} \\
\midrule
\endhead
\midrule\multicolumn{6}{r@{}}{\textit{Continued on next page}}\\
\endfoot
\bottomrule
\endlastfoot
Terminal & Quality & 50 & 445 & 51 & 546 \\
Terminal & Cost & 235 & 45 & 233 & 513 \\
\addlinespace
HumanEvalFix & Quality & 47 & 1158 & 49 & 1254 \\
HumanEvalFix & Cost & 827 & 41 & 324 & 1192 \\
\addlinespace
Omni100 & Quality & 6 & 176 & 3 & 185 \\
Omni100 & Cost & 73 & 15 & 97 & 185 \\
\addlinespace
Omni300 & Quality & 6 & 607 & 12 & 625 \\
Omni300 & Cost & 186 & 57 & 435 & 678 \\
\end{xltabular}
\endgroup

%% file: appendix/appendix_guidance.tex
\begin{appendixtext}
\section{Supplementary Results for Simulated Roles and Guidance Behaviors}\label{app:guidance}

Common support for role comparisons is determined by the root task and the contracts for the source execution and continuations. For HumanEvalFix, both source executions and continuations use GLM-4.7-Flash \citep{zai2026glm47flash,glm2025glm45}, with guidance from DeepSeek-V4-Flash \citep{deepseek2026v4}. For Terminal, source executions use GLM-4.7-Flash, archived guidance uses DeepSeek-V4-Flash, and continuations use DeepSeek-V4-Flash. The contracts retain the model, budget, environment, prompt, and acceptance fields; the role prompt is the prespecified comparison variable. Roles may differ in their number of turns and intervention timing, so role comparisons use aggregate proportions over common tasks.

\subsection{Linguistic Functions and Information Changes}
Tables~\ref{tab:semanticsfull} and~\ref{tab:semanticsfullinc} report the primary-function and information-increment distributions for each simulated role. Information increments can co-occur and use their own valid denominator. Table~\ref{tab:repeatcontent} cross-tabulates repetition and substantive information increments: repetition can accompany new information, refinement, or correction.
\end{appendixtext}

\begingroup\small
\setlength{\LTcapwidth}{.70\textwidth}
\begin{xltabular}{.70\textwidth}{@{}>{\raggedright\arraybackslash}Xrr@{}}
\caption{Complete primary-function statistics for simulated roles. Primary function is a single-label dimension; information increments may co-occur and use a different valid denominator.}\label{tab:semanticsfull}\\
\toprule
\textbf{Primary function} & \textbf{\shortstack{Expert-style\\$n$ (\%)}} & \textbf{\shortstack{Ordinary-style\\$n$ (\%)}} \\
\midrule
\endfirsthead
\multicolumn{3}{@{}l}{Table~\thetable\ (continued)}\\
\toprule
\textbf{Primary function} & \textbf{\shortstack{Expert-style\\$n$ (\%)}} & \textbf{\shortstack{Ordinary-style\\$n$ (\%)}} \\
\midrule
\endhead
\midrule\multicolumn{3}{r@{}}{\textit{Continued on next page}}\\
\endfoot
\bottomrule
\endlastfoot
Implementation repair & 674 (36.12) & 655 (31.87) \\
Acceptance & 203 (10.88) & 170 (8.27) \\
Strategy guidance & 92 (4.93) & 62 (3.02) \\
Constraints & 16 (0.86) & 9 (0.44) \\
Checking and verification & 574 (30.76) & 626 (30.46) \\
Other or unclear & 1 (0.05) & 0 (0.00) \\
Information provision & 1 (0.05) & 1 (0.05) \\
Social expression & 0 (0.00) & 0 (0.00) \\
Process control & 76 (4.07) & 111 (5.40) \\
New task & 180 (9.65) & 236 (11.48) \\
Problem diagnosis & 37 (1.98) & 91 (4.43) \\
Explanation and clarification & 12 (0.64) & 94 (4.57) \\
\end{xltabular}
\endgroup

\begingroup\small
\setlength{\LTcapwidth}{.70\textwidth}
\begin{xltabular}{.70\textwidth}{@{}>{\raggedright\arraybackslash}Xrr@{}}
\caption{Complete information-increment statistics for simulated roles. Information increments may co-occur and use a different valid denominator from primary functions.}\label{tab:semanticsfullinc}\\
\toprule
\textbf{Information increment} & \textbf{\shortstack{Expert-style\\$n$ (\%)}} & \textbf{\shortstack{Ordinary-style\\$n$ (\%)}} \\
\midrule
\endfirsthead
\multicolumn{3}{@{}l}{Table~\thetable\ (continued)}\\
\toprule
\textbf{Information increment} & \textbf{\shortstack{Expert-style\\$n$ (\%)}} & \textbf{\shortstack{Ordinary-style\\$n$ (\%)}} \\
\midrule
\endhead
\midrule\multicolumn{3}{r@{}}{\textit{Continued on next page}}\\
\endfoot
\bottomrule
\endlastfoot
New information & 178 (15.74) & 96 (7.61) \\
Refinement & 923 (81.61) & 949 (75.26) \\
Correction & 464 (41.03) & 449 (35.61) \\
Change of approach & 163 (14.41) & 133 (10.55) \\
Acknowledgment & 444 (39.26) & 454 (36.00) \\
New requirement & 195 (17.24) & 203 (16.10) \\
Unclear & 1 (0.09) & 1 (0.08) \\
Repetition & 638 (56.41) & 767 (60.82) \\
\end{xltabular}
\endgroup

\begingroup\small
\setlength{\LTcapwidth}{.70\textwidth}
\begin{xltabular}{.70\textwidth}{@{}>{\raggedright\arraybackslash}Xrr@{}}
\caption{Role distributions of repetition and substantive information. Repetition can co-occur with new information, refinement, or correction; a repetition label alone does not establish an absence of contribution.}\label{tab:repeatcontent}\\
\toprule
\textbf{\shortstack{Repetition and\\increment}} & \textbf{\shortstack{Expert-style\\$n$ (\%)}} & \textbf{\shortstack{Ordinary-style\\$n$ (\%)}} \\
\midrule
\endfirsthead
\multicolumn{3}{@{}l}{Table~\thetable\ (continued)}\\
\toprule
\textbf{\shortstack{Repetition and\\increment}} & \textbf{\shortstack{Expert-style\\$n$ (\%)}} & \textbf{\shortstack{Ordinary-style\\$n$ (\%)}} \\
\midrule
\endhead
\midrule\multicolumn{3}{r@{}}{\textit{Continued on next page}}\\
\endfoot
\bottomrule
\endlastfoot
Repetition, with increment & 609 (53.85) & 670 (53.13) \\
Repetition, no increment & 29 (2.56) & 97 (7.69) \\
No repetition, with increment & 355 (31.39) & 364 (28.87) \\
No repetition, no increment & 138 (12.20) & 130 (10.31) \\
Total & 1,131 (100.00) & 1,261 (100.00) \\
\end{xltabular}
\endgroup

\begin{appendixtext}
\subsection{Selected Local Computation Terms}
Table~\ref{tab:localledger} summarizes selected local computation terms for Terminal and Omni100.
\end{appendixtext}

\begingroup\small
\begin{xltabular}{\textwidth}{@{}>{\raggedright\arraybackslash}Xlrrrr@{}}
\caption{Archived local computation terms by role. This table describes the structure of selected functions; it does not constitute an exhaustive turn-level role comparison under the current common contract.}\label{tab:localledger}\\
\toprule
\textbf{Setting} & \textbf{Role} & \textbf{\shortstack{Source turns\\$N$}} & \textbf{\shortstack{Repair\\(k)}} & \textbf{\shortstack{Refinement\\(k)}} & \textbf{\shortstack{Acknowledgment\\(k)}} \\
\midrule
\endfirsthead
\multicolumn{6}{@{}l}{Table~\thetable\ (continued)}\\
\toprule
\textbf{Setting} & \textbf{Role} & \textbf{\shortstack{Source turns\\$N$}} & \textbf{\shortstack{Repair\\(k)}} & \textbf{\shortstack{Refinement\\(k)}} & \textbf{\shortstack{Acknowledgment\\(k)}} \\
\midrule
\endhead
\midrule\multicolumn{6}{r@{}}{\textit{Continued on next page}}\\
\endfoot
\bottomrule
\endlastfoot
Terminal & Expert-style & 269 & +0.95 & +5.00 & -3.21 \\
Terminal & Ordinary-style & 286 & +12.07 & +12.90 & +1.64 \\
Omni100 & Expert-style & 109 & +1.58 & +1.37 & -0.15 \\
Omni100 & Ordinary-style & 153 & +1.00 & +2.06 & -0.20 \\
\end{xltabular}
\endgroup

\begin{appendixtext}
\subsection{Terminal Outcomes and Local Ledger}
Table~\ref{tab:terminalcard} reports full-session task quality and token use for Terminal. Table~\ref{tab:terminalcard_ledger} gives the corresponding category-level local computation ledger.
\end{appendixtext}

\begingroup\small
\setlength{\LTcapwidth}{.70\textwidth}
\begin{xltabular}{.70\textwidth}{@{}>{\raggedright\arraybackslash}Xrr@{}}
\caption{Full-session results for simulated roles in Terminal. Common task support for quality and full-session cost is 15 and 16 tasks, respectively.}\label{tab:terminalcard}\\
\toprule
\textbf{Measure} & \textbf{Expert-style} & \textbf{Ordinary-style} \\
\midrule
\endfirsthead
\multicolumn{3}{@{}l}{Table~\thetable\ (continued)}\\
\toprule
\textbf{Measure} & \textbf{Expert-style} & \textbf{Ordinary-style} \\
\midrule
\endhead
\midrule\multicolumn{3}{r@{}}{\textit{Continued on next page}}\\
\endfoot
\bottomrule
\endlastfoot
Task quality (\%) & 81.04 & 70.65 \\
Full-session tokens / task & 336,016.19 & 350,128.31 \\
\end{xltabular}
\endgroup

\begingroup\small
\setlength{\LTcapwidth}{.80\textwidth}
\begin{xltabular}{.80\textwidth}{@{}>{\raggedright\arraybackslash}Xrrrr@{}}
\caption{Terminal guidance behaviors and local computation ledger. $C$ is category proportion multiplied by the category's mean change in remaining tokens (k); positive values indicate additional computation and negative values indicate savings. $n$ (\%) is the category count and its share of source turns within the role. $\dagger$ marks fewer than 10 cost-valid turns; NA denotes no cost record; --- denotes an undefined value. Information increments may co-occur: neither their counts nor $C$ terms sum to unique turns or decompose full-session cost. Primary-function $C$ terms also cannot reconstruct the total of a separately frozen full-session cohort.}\label{tab:terminalcard_ledger}\\
\toprule
 & \multicolumn{2}{c}{\textbf{Expert-style}} & \multicolumn{2}{c}{\textbf{Ordinary-style}} \\
\cmidrule(lr){2-3}\cmidrule(lr){4-5}
\textbf{Category} & \textbf{$n$ (\%)} & \textbf{$C$ (k)} & \textbf{$n$ (\%)} & \textbf{$C$ (k)} \\
\midrule
\endfirsthead
\multicolumn{5}{@{}l}{Table~\thetable\ (continued)}\\
\toprule
 & \multicolumn{2}{c}{\textbf{Expert-style}} & \multicolumn{2}{c}{\textbf{Ordinary-style}} \\
\cmidrule(lr){2-3}\cmidrule(lr){4-5}
\textbf{Category} & \textbf{$n$ (\%)} & \textbf{$C$ (k)} & \textbf{$n$ (\%)} & \textbf{$C$ (k)} \\
\midrule
\endhead
\midrule\multicolumn{5}{r@{}}{\textit{Continued on next page}}\\
\endfoot
\bottomrule
\endlastfoot
\multicolumn{5}{@{}l}{\textit{Primary function}} \\
Process control & 24 (8.9) & -6.92\ensuremath{\dagger} & 30 (10.5) & -9.91\ensuremath{\dagger} \\
Strategy guidance & 32 (11.9) & +2.68 & 17 (5.9) & -0.47\ensuremath{\dagger} \\
Acceptance & 0 (0.0) & --- & 1 (0.3) & -0.03\ensuremath{\dagger} \\
Constraints & 6 (2.2) & +0.11\ensuremath{\dagger} & 2 (0.7) & +0.83\ensuremath{\dagger} \\
Explanation and clarification & 4 (1.5) & +0.61\ensuremath{\dagger} & 11 (3.8) & +2.47\ensuremath{\dagger} \\
Problem diagnosis & 7 (2.6) & +0.77\ensuremath{\dagger} & 22 (7.7) & +1.11 \\
Implementation repair & 82 (30.5) & +0.95 & 75 (26.2) & +12.07 \\
Checking and verification & 62 (23.0) & +2.44 & 58 (20.3) & +6.71 \\
New task & 52 (19.3) & +3.39 & 69 (24.1) & +8.13 \\
Information provision & 0 (0.0) & --- & 1 (0.3) & NA \\
\addlinespace\multicolumn{5}{@{}l}{\textit{Information increment}} \\
Acknowledgment & 16 (5.9) & -3.21 & 20 (7.0) & +1.64 \\
Change of approach & 38 (14.1) & -1.52 & 46 (16.1) & -1.93 \\
New information & 18 (6.7) & +1.27 & 6 (2.1) & +0.06\ensuremath{\dagger} \\
New requirement & 26 (9.7) & +1.48 & 21 (7.3) & +0.82 \\
Correction & 68 (25.3) & +3.16 & 59 (20.6) & +6.91 \\
Repetition & 89 (33.1) & +4.10 & 91 (31.8) & +10.52 \\
Refinement & 130 (48.3) & +5.00 & 121 (42.3) & +12.90 \\
Annotation not accepted & 138 (51.3) & NA & 156 (54.5) & NA \\
Source turns & 269 & --- & 286 & --- \\
\end{xltabular}
\endgroup

\begin{appendixtext}
\subsection{Omni100 Outcomes and Local Ledger}
Table~\ref{tab:omnicard} reports full-session task quality and token use for Omni100. Table~\ref{tab:omnicard_ledger} gives the corresponding category-level local computation ledger.
\end{appendixtext}

\begingroup\small
\setlength{\LTcapwidth}{.70\textwidth}
\begin{xltabular}{.70\textwidth}{@{}>{\raggedright\arraybackslash}Xrr@{}}
\caption{Full-session results for simulated roles in Omni100. Common task support for quality and full-session cost is 19 and 19 tasks, respectively.}\label{tab:omnicard}\\
\toprule
\textbf{Measure} & \textbf{Expert-style} & \textbf{Ordinary-style} \\
\midrule
\endfirsthead
\multicolumn{3}{@{}l}{Table~\thetable\ (continued)}\\
\toprule
\textbf{Measure} & \textbf{Expert-style} & \textbf{Ordinary-style} \\
\midrule
\endhead
\midrule\multicolumn{3}{r@{}}{\textit{Continued on next page}}\\
\endfoot
\bottomrule
\endlastfoot
Task quality (\%) & 100.00 & 100.00 \\
Full-session tokens / task & 9,309.26 & 10,414.95 \\
\end{xltabular}
\endgroup

\begingroup\small
\setlength{\LTcapwidth}{.80\textwidth}
\begin{xltabular}{.80\textwidth}{@{}>{\raggedright\arraybackslash}Xrrrr@{}}
\caption{Omni100 guidance behaviors and local computation ledger. $C$ is category proportion multiplied by the category's mean change in remaining tokens (k); positive values indicate additional computation and negative values indicate savings. $n$ (\%) is the category count and its share of source turns within the role. $\dagger$ marks fewer than 10 cost-valid turns; NA denotes no cost record; --- denotes an undefined value. Information increments may co-occur: neither their counts nor $C$ terms sum to unique turns or decompose full-session cost. Primary-function $C$ terms also cannot reconstruct the total of a separately frozen full-session cohort.}\label{tab:omnicard_ledger}\\
\toprule
 & \multicolumn{2}{c}{\textbf{Expert-style}} & \multicolumn{2}{c}{\textbf{Ordinary-style}} \\
\cmidrule(lr){2-3}\cmidrule(lr){4-5}
\textbf{Category} & \textbf{$n$ (\%)} & \textbf{$C$ (k)} & \textbf{$n$ (\%)} & \textbf{$C$ (k)} \\
\midrule
\endfirsthead
\multicolumn{5}{@{}l}{Table~\thetable\ (continued)}\\
\toprule
 & \multicolumn{2}{c}{\textbf{Expert-style}} & \multicolumn{2}{c}{\textbf{Ordinary-style}} \\
\cmidrule(lr){2-3}\cmidrule(lr){4-5}
\textbf{Category} & \textbf{$n$ (\%)} & \textbf{$C$ (k)} & \textbf{$n$ (\%)} & \textbf{$C$ (k)} \\
\midrule
\endhead
\midrule\multicolumn{5}{r@{}}{\textit{Continued on next page}}\\
\endfoot
\bottomrule
\endlastfoot
\multicolumn{5}{@{}l}{\textit{Primary function}} \\
Strategy guidance & 6 (5.5) & -0.09\ensuremath{\dagger} & 6 (3.9) & -0.08\ensuremath{\dagger} \\
Acceptance & 11 (10.1) & -0.06\ensuremath{\dagger} & 14 (9.2) & -0.08\ensuremath{\dagger} \\
Explanation and clarification & 0 (0.0) & --- & 6 (3.9) & -0.13\ensuremath{\dagger} \\
Problem diagnosis & 8 (7.3) & -0.08\ensuremath{\dagger} & 15 (9.8) & +0.44\ensuremath{\dagger} \\
Implementation repair & 64 (58.7) & +1.58 & 70 (45.8) & +1.00 \\
Checking and verification & 16 (14.7) & +0.06\ensuremath{\dagger} & 38 (24.8) & +0.74 \\
New task & 4 (3.7) & -0.12\ensuremath{\dagger} & 4 (2.6) & +0.19\ensuremath{\dagger} \\
\addlinespace\multicolumn{5}{@{}l}{\textit{Information increment}} \\
Acknowledgment & 32 (29.4) & -0.15 & 22 (14.4) & -0.20 \\
Change of approach & 17 (15.6) & -0.17 & 11 (7.2) & +0.02\ensuremath{\dagger} \\
New information & 59 (54.1) & -0.16 & 32 (20.9) & +0.87 \\
New requirement & 12 (11.0) & +0.68\ensuremath{\dagger} & 26 (17.0) & +0.40 \\
Correction & 83 (76.1) & +1.75 & 117 (76.5) & +1.77 \\
Repetition & 66 (60.6) & +0.13 & 92 (60.1) & +1.38 \\
Refinement & 97 (89.0) & +1.37 & 136 (88.9) & +2.06 \\
Unclear & 1 (0.9) & +0.003\ensuremath{\dagger} & 0 (0.0) & --- \\
Source turns & 109 & --- & 153 & --- \\
\end{xltabular}
\endgroup

%% file: appendix/appendix_naturalness.tex
\begin{appendixtext}
% Keep these consecutive prose sections in the same column flow.
\input{appendix/appendix_human}

\section{Blinded Naturalness Assessment of Simulated Utterances}\label{app:naturalness}

The naturalness study compares original human utterances, prompt-based simulations, and utterances from Qwen2.5-7B-Instruct \citep{yang2024qwen25} adapted using approximately 17k WildChat/DevGPT exchanges. Ten evaluators with experience in agent programming each complete 30 choices; the archive contains 300 judgments and 285 valid choices. The analysis further stratifies exchanges by whether an exact match occurs in the adaptation training or validation data, keeping samples with overlap distinct from the remainder.

Table~\ref{tab:naturalness} reports the subset with no exact matching exchange. Across its 131 valid choices, original human utterances, prompt-based simulations, and the dialogue-adapted model receive 32.1\%, 26.7\%, and 41.2\%, respectively. The distributions differ between code and text scenarios, so their naturalness results are reported separately. The measure is the frequency with which a candidate utterance is judged more human-like, providing a check of simulated language form. Human-origin task effects are assessed separately through execution outcomes.
\end{appendixtext}

\begingroup\small
\setlength{\LTcapwidth}{.70\textwidth}
\begin{xltabular}{.70\textwidth}{@{}>{\raggedright\arraybackslash}Xrrr@{}}
\caption{Naturalness choice rates (\%) in the subset with no exact matching exchange found. $N$ is the number of valid choices; three candidate sources compete within each judgment.}\label{tab:naturalness}\\
\toprule
\textbf{Candidate source} & \textbf{\shortstack{Code\\$N=47$}} & \textbf{\shortstack{Text\\$N=84$}} & \textbf{\shortstack{Total\\$N=131$}} \\
\midrule
\endfirsthead
\multicolumn{4}{@{}l}{Table~\thetable\ (continued)}\\
\toprule
\textbf{Candidate source} & \textbf{\shortstack{Code\\$N=47$}} & \textbf{\shortstack{Text\\$N=84$}} & \textbf{\shortstack{Total\\$N=131$}} \\
\midrule
\endhead
\midrule\multicolumn{4}{r@{}}{\textit{Continued on next page}}\\
\endfoot
\bottomrule
\endlastfoot
Original human & 29.8 & 33.3 & 32.1 \\
Prompt-based simulation & 34.0 & 22.6 & 26.7 \\
Dialogue-adapted 7B & 36.2 & 44.0 & 41.2 \\
\end{xltabular}
\endgroup

\begin{appendixtext}
Table~\ref{tab:naturalnessall} reports all valid choices and the overlap-stratified summaries. These judgment counts are separate from the HumanBench task-comparison denominators.
\end{appendixtext}

\begingroup\small
\begin{xltabular}{\textwidth}{@{}>{\raggedright\arraybackslash}Xrrrrrr@{}}
\caption{All valid choices and overlap-stratified summaries from the original naturalness study. Its judgment counts, scenario subsets, and candidate sources are separate from the denominators for HumanBench task comparisons.}\label{tab:naturalnessall}\\
\toprule
\textbf{Evaluation sample} & \textbf{Items} & \textbf{\shortstack{Submitted\\/ valid}} & \textbf{Abstentions} & \textbf{\shortstack{Human\\(\%)}} & \textbf{\shortstack{Prompt\\(\%)}} & \textbf{\shortstack{7B\\(\%)}} \\
\midrule
\endfirsthead
\multicolumn{7}{@{}l}{Table~\thetable\ (continued)}\\
\toprule
\textbf{Evaluation sample} & \textbf{Items} & \textbf{\shortstack{Submitted\\/ valid}} & \textbf{Abstentions} & \textbf{\shortstack{Human\\(\%)}} & \textbf{\shortstack{Prompt\\(\%)}} & \textbf{\shortstack{7B\\(\%)}} \\
\midrule
\endhead
\midrule\multicolumn{7}{r@{}}{\textit{Continued on next page}}\\
\endfoot
\bottomrule
\endlastfoot
All judgments & 229 & 300/285 & 15 & 34.7 & 26.3 & 38.9 \\
Exact training match & 117 & 131/121 & 10 & 40.5 & 24.0 & 35.5 \\
Exact validation match & 22 & 35/33 & 2 & 24.2 & 33.3 & 42.4 \\
No exact match found & 90 & 134/131 & 3 & 32.1 & 26.7 & 41.2 \\
Successful candidates: all & 217 & 288/274 & 14 & 34.7 & 25.2 & 40.1 \\
Successful candidates: no exact match & 88 & 132/129 & 3 & 32.6 & 25.6 & 41.9 \\
\end{xltabular}
\endgroup

%% file: appendix/appendix_human.tex
\section{HumanBench Data Provenance}
\label{app:humanbench-provenance}

HumanBench contributes 85 human-origin source sessions covering 49 root tasks and 600 adjacent checkpoint pairs. Of these pairs, 506 have valid task-quality observations, complete token records, and matched continuation contracts. The 85 source sessions are not 85 independent participants. The paper reports only aggregate execution-based measurements and does not distribute original messages, code, screenshots, or model-assisted semantic-category analyses.

%% file: main.bbl
\begin{thebibliography}{39}
\providecommand{\natexlab}[1]{#1}

\bibitem[{Acemoglu and Restrepo(2019)}]{acemoglu2019tasks}
Daron Acemoglu and Pascual Restrepo. 2019.
\newblock \href {https://doi.org/10.1257/jep.33.2.3} {Automation and new tasks: How technology displaces and reinstates labor}.
\newblock \emph{Journal of Economic Perspectives}, 33(2):3--30.

\bibitem[{Amershi et~al.(2019)Amershi, Weld, Vorvoreanu, Fourney, Nushi, Collisson, Suh, Iqbal, Bennett, Inkpen, Teevan, Kikin-Gil, and Horvitz}]{amershi2019guidelines}
Saleema Amershi, Dan Weld, Mihaela Vorvoreanu, Adam Fourney, Besmira Nushi, Penny Collisson, Jina Suh, Shamsi Iqbal, Paul~N. Bennett, Kori Inkpen, Jaime Teevan, Ruth Kikin-Gil, and Eric Horvitz. 2019.
\newblock \href {https://doi.org/10.1145/3290605.3300233} {{Guidelines for Human-AI Interaction}}.
\newblock In \emph{Proceedings of the 2019 CHI Conference on Human Factors in Computing Systems}.

\bibitem[{Chang et~al.(2025)Chang, Anderson, and Hofman}]{chang2025chatbench}
Serina Chang, Ashton Anderson, and Jake~M. Hofman. 2025.
\newblock \href {https://doi.org/10.18653/v1/2025.acl-long.1262} {{ChatBench}: From static benchmarks to human-{AI} evaluation}.
\newblock In \emph{Proceedings of the 63rd Annual Meeting of the Association for Computational Linguistics (Volume 1: Long Papers)}, pages 26009--26038. Association for Computational Linguistics.

\bibitem[{{DeepSeek-AI}(2026)}]{deepseek2026v4}
{DeepSeek-AI}. 2026.
\newblock \href {https://arxiv.org/abs/2606.19348} {{DeepSeek-V4}: Towards highly efficient million-token context intelligence}.
\newblock \emph{Preprint}, arXiv:2606.19348.

\bibitem[{Dou et~al.(2025)Dou, Galley, Peng, Kedzie, Cai, Ritter, Quirk, Xu, and Gao}]{dou2025simulatorarena}
Yao Dou, Michel Galley, Baolin Peng, Chris Kedzie, Weixin Cai, Alan Ritter, Chris Quirk, Wei Xu, and Jianfeng Gao. 2025.
\newblock \href {https://doi.org/10.18653/v1/2025.emnlp-main.1786} {{SimulatorArena}: Are user simulators reliable proxies for multi-turn evaluation of {AI} assistants?}
\newblock In \emph{Proceedings of the 2025 Conference on Empirical Methods in Natural Language Processing}, pages 35212--35290. Association for Computational Linguistics.

\bibitem[{Gao et~al.(2025)Gao, Song, Yang, Cai, Miao, Dong, Li, Ma, Chen, Xu, Tang, Wang, Zan, Quan, Zhang, Sha, Zhang, Ren, Liu, and Chang}]{gao2025omnimath}
Bofei Gao, Feifan Song, Zhe Yang, Zefan Cai, Yibo Miao, Qingxiu Dong, Lei Li, Chenghao Ma, Liang Chen, Runxin Xu, Zhengyang Tang, Benyou Wang, Daoguang Zan, Shanghaoran Quan, Ge~Zhang, Lei Sha, Yichang Zhang, Xuancheng Ren, Tianyu Liu, and Baobao Chang. 2025.
\newblock \href {https://proceedings.iclr.cc/paper_files/paper/2025/hash/f9e1e8b56c7e363985ebeb0e9dd1a85c-Abstract-Conference.html} {{Omni-MATH}: A universal olympiad level mathematic benchmark for large language models}.
\newblock In \emph{The Thirteenth International Conference on Learning Representations}.

\bibitem[{{GLM-4.5 Team} et~al.(2025){GLM-4.5 Team}, Zeng, Lv, Zheng et~al.}]{glm2025glm45}
{GLM-4.5 Team}, Aohan Zeng, Xin Lv, Qinkai Zheng, et~al. 2025.
\newblock \href {https://arxiv.org/abs/2508.06471} {{GLM-4.5}: Agentic, reasoning, and coding ({ARC}) foundation models}.
\newblock \emph{Preprint}, arXiv:2508.06471.

\bibitem[{Gunjal et~al.(2026)Gunjal, Wang, Lau, Nath, He, Liu, and Hendryx}]{gunjal2026rar}
Anisha Gunjal, Anthony Wang, Elaine Lau, Vaskar Nath, Yunzhong He, Bing Liu, and Sean Hendryx. 2026.
\newblock \href {https://proceedings.iclr.cc/paper_files/paper/2026/hash/cfd7bee7a651ee9af525098ef67a9e45-Abstract-Conference.html} {Rubrics as rewards: Reinforcement learning beyond verifiable domains}.
\newblock In \emph{The Fourteenth International Conference on Learning Representations}.

\bibitem[{Hashemi et~al.(2024)Hashemi, Eisner, Rosset, Van~Durme, and Kedzie}]{hashemi2024llmrubric}
Helia Hashemi, Jason Eisner, Corby Rosset, Benjamin Van~Durme, and Chris Kedzie. 2024.
\newblock \href {https://doi.org/10.18653/v1/2024.acl-long.745} {{LLM}-rubric: A multidimensional, calibrated approach to automated evaluation of natural language texts}.
\newblock In \emph{Proceedings of the 62nd Annual Meeting of the Association for Computational Linguistics (Volume 1: Long Papers)}, pages 13806--13834. Association for Computational Linguistics.

\bibitem[{Horvitz(1999)}]{horvitz1999mixedinitiative}
Eric Horvitz. 1999.
\newblock \href {https://doi.org/10.1145/302979.303030} {{Principles of Mixed-Initiative User Interfaces}}.
\newblock In \emph{Proceedings of the SIGCHI Conference on Human Factors in Computing Systems}, pages 159--166.

\bibitem[{Jimenez et~al.(2024)Jimenez, Yang, Wettig, Yao, Pei, Press, and Narasimhan}]{jimenez2024swebench}
Carlos~E. Jimenez, John Yang, Alexander Wettig, Shunyu Yao, Kexin Pei, Ofir Press, and Karthik Narasimhan. 2024.
\newblock \href {https://proceedings.iclr.cc/paper_files/paper/2024/hash/edac78c3e300629acfe6cbe9ca88fb84-Abstract-Conference.html} {{SWE-bench}: Can language models resolve real-world {GitHub} issues?}
\newblock In \emph{The Twelfth International Conference on Learning Representations}.

\bibitem[{Kapoor et~al.(2024)Kapoor, Stroebl, Siegel, Nadgir, and Narayanan}]{kapoor2024agents}
Sayash Kapoor, Benedikt Stroebl, Zachary~S. Siegel, Nitya Nadgir, and Arvind Narayanan. 2024.
\newblock \href {https://arxiv.org/abs/2407.01502} {{AI} agents that matter}.
\newblock \emph{Preprint}, arXiv:2407.01502.

\bibitem[{Kim et~al.(2024)Kim, Suk, Longpre, Lin, Shin, Welleck, Neubig, Lee, Lee, and Seo}]{kim2024prometheus2}
Seungone Kim, Juyoung Suk, Shayne Longpre, Bill~Yuchen Lin, Jamin Shin, Sean Welleck, Graham Neubig, Moontae Lee, Kyungjae Lee, and Minjoon Seo. 2024.
\newblock \href {https://doi.org/10.18653/v1/2024.emnlp-main.248} {{Prometheus 2: An Open Source Language Model Specialized in Evaluating Other Language Models}}.
\newblock In \emph{Proceedings of the 2024 Conference on Empirical Methods in Natural Language Processing}, pages 4334--4353.

\bibitem[{Laban et~al.(2026)Laban, Hayashi, Zhou, and Neville}]{laban2026lost}
Philippe Laban, Hiroaki Hayashi, Yingbo Zhou, and Jennifer Neville. 2026.
\newblock \href {https://openreview.net/forum?id=VKGTGGcwl6} {{LLMs Get Lost In Multi-Turn Conversation}}.
\newblock In \emph{The Fourteenth International Conference on Learning Representations}.

\bibitem[{Lee et~al.(2022)Lee, Liang, and Yang}]{lee2022coauthor}
Mina Lee, Percy Liang, and Qian Yang. 2022.
\newblock \href {https://doi.org/10.1145/3491102.3502030} {{CoAuthor: Designing a Human-AI Collaborative Writing Dataset for Exploring Language Model Capabilities}}.
\newblock In \emph{Proceedings of the 2022 CHI Conference on Human Factors in Computing Systems}.

\bibitem[{Lightman et~al.(2024)Lightman, Kosaraju, Burda, Edwards, Baker, Lee, Leike, Schulman, Sutskever, and Cobbe}]{lightman2024verify}
Hunter Lightman, Vineet Kosaraju, Yuri Burda, Harrison Edwards, Bowen Baker, Teddy Lee, Jan Leike, John Schulman, Ilya Sutskever, and Karl Cobbe. 2024.
\newblock \href {https://proceedings.iclr.cc/paper_files/paper/2024/hash/aca97732e30bcf1303bc22ac3924fd16-Abstract-Conference.html} {{Let's Verify Step by Step}}.
\newblock In \emph{The Twelfth International Conference on Learning Representations}.

\bibitem[{Liu et~al.(2024)Liu, Yu, Zhang, Xu, Lei, Lai, Gu, Ding, Men, Yang, Zhang, Deng, Zeng, Du, Zhang, Shen, Zhang, Su, Sun, Huang, Dong, and Tang}]{liu2024agentbench}
Xiao Liu, Hao Yu, Hanchen Zhang, Yifan Xu, Xuanyu Lei, Hanyu Lai, Yu~Gu, Hangliang Ding, Kaiwen Men, Kejuan Yang, Shudan Zhang, Xiang Deng, Aohan Zeng, Zhengxiao Du, Chenhui Zhang, Sheng Shen, Tianjun Zhang, Yu~Su, Huan Sun, and 3 others. 2024.
\newblock \href {https://arxiv.org/abs/2308.03688} {{AgentBench: Evaluating LLMs as Agents}}.
\newblock In \emph{The Twelfth International Conference on Learning Representations}.

\bibitem[{Liu et~al.(2023)Liu, Iter, Xu, Wang, Xu, and Zhu}]{liu2023geval}
Yang Liu, Dan Iter, Yichong Xu, Shuohang Wang, Ruochen Xu, and Chenguang Zhu. 2023.
\newblock \href {https://doi.org/10.18653/v1/2023.emnlp-main.153} {{G-Eval: NLG Evaluation using Gpt-4 with Better Human Alignment}}.
\newblock In \emph{Proceedings of the 2023 Conference on Empirical Methods in Natural Language Processing}, pages 2511--2522.

\bibitem[{Ma et~al.(2024)Ma, Zhang, Zhu, Yang, Yang, Jin, Lan, Kong, and He}]{ma2024agentboard}
Chang Ma, Junlei Zhang, Zhihao Zhu, Cheng Yang, Yujiu Yang, Yaohui Jin, Zhenzhong Lan, Lingpeng Kong, and Junxian He. 2024.
\newblock \href {https://doi.org/10.52202/079017-2365} {{AgentBoard}: An analytical evaluation board of multi-turn {LLM} agents}.
\newblock In \emph{Advances in Neural Information Processing Systems}, volume~37.

\bibitem[{Madaan et~al.(2023)Madaan, Tandon, Gupta, Hallinan, Gao, Wiegreffe, Alon, Dziri, Prabhumoye, Yang, Gupta, Majumder, Hermann, Welleck, Yazdanbakhsh, and Clark}]{madaan2023selfrefine}
Aman Madaan, Niket Tandon, Prakhar Gupta, Skyler Hallinan, Luyu Gao, Sarah Wiegreffe, Uri Alon, Nouha Dziri, Shrimai Prabhumoye, Yiming Yang, Shashank Gupta, Bodhisattwa~Prasad Majumder, Katherine Hermann, Sean Welleck, Amir Yazdanbakhsh, and Peter Clark. 2023.
\newblock \href {https://proceedings.neurips.cc/paper_files/paper/2023/hash/91edff07232fb1b55a505a9e9f6c0ff3-Abstract-Conference.html} {{Self-Refine: Iterative Refinement with Self-Feedback}}.
\newblock In \emph{Advances in Neural Information Processing Systems}, volume~36.

\bibitem[{Merrill et~al.(2026)Merrill, Shaw, Carlini et~al.}]{merrill2026terminalbench}
Mike~A. Merrill, Alexander~G. Shaw, Nicholas Carlini, et~al. 2026.
\newblock \href {https://arxiv.org/abs/2601.11868} {{Terminal-Bench}: Benchmarking agents on hard, realistic tasks in command line interfaces}.
\newblock \emph{Preprint}, arXiv:2601.11868.

\bibitem[{Muennighoff et~al.(2024)Muennighoff, Liu, Zebaze, Zheng, Hui, Zhuo, Singh, Tang, von Werra, and Longpre}]{muennighoff2024octopack}
Niklas Muennighoff, Qian Liu, Armel Zebaze, Qinkai Zheng, Binyuan Hui, Terry~Yue Zhuo, Swayam Singh, Xiangru Tang, Leandro von Werra, and Shayne Longpre. 2024.
\newblock \href {https://proceedings.iclr.cc/paper_files/paper/2024/hash/1ec299a5229034141e58aeded0d0b9de-Abstract-Conference.html} {{OctoPack}: Instruction tuning code large language models}.
\newblock In \emph{The Twelfth International Conference on Learning Representations}.

\bibitem[{Shao et~al.(2026{\natexlab{a}})Shao, Samuel, Jiang, Yang, and Yang}]{shao2026collaborativegym}
Yijia Shao, Vinay Samuel, Yucheng Jiang, John Yang, and Diyi Yang. 2026{\natexlab{a}}.
\newblock \href {https://openreview.net/forum?id=GDYueXtKXT} {Collaborative gym: A framework for enabling and evaluating human-agent collaboration}.
\newblock In \emph{The Fourteenth International Conference on Learning Representations}.

\bibitem[{Shao et~al.(2026{\natexlab{b}})Shao, Wang, Ahuja, Wang, Liu, and Yang}]{shao2026collabskill}
Yijia Shao, Zora~Zhiruo Wang, Neel Ahuja, Yicheng Wang, Bowen Liu, and Diyi Yang. 2026{\natexlab{b}}.
\newblock \href {https://arxiv.org/abs/2606.09833} {{CollabSkill}: Evaluating human-agent collaboration on real-world tasks}.
\newblock \emph{Preprint}, arXiv:2606.09833.

\bibitem[{Shinn et~al.(2023)Shinn, Cassano, Gopinath, Narasimhan, and Yao}]{shinn2023reflexion}
Noah Shinn, Federico Cassano, Ashwin Gopinath, Karthik Narasimhan, and Shunyu Yao. 2023.
\newblock \href {https://proceedings.neurips.cc/paper_files/paper/2023/hash/1b44b878bb782e6954cd888628510e90-Abstract-Conference.html} {{Reflexion: Language Agents with Verbal Reinforcement Learning}}.
\newblock In \emph{Advances in Neural Information Processing Systems}, volume~36.

\bibitem[{Snell et~al.(2025)Snell, Lee, Xu, and Kumar}]{snell2025compute}
Charlie Snell, Jaehoon Lee, Kelvin Xu, and Aviral Kumar. 2025.
\newblock \href {https://proceedings.iclr.cc/paper_files/paper/2025/hash/1b623663fd9b874366f3ce019fdfdd44-Abstract-Conference.html} {{Scaling LLM Test-Time Compute Optimally Can be More Effective than Scaling Parameters for Reasoning}}.
\newblock In \emph{The Thirteenth International Conference on Learning Representations}.

\bibitem[{Vaccaro et~al.(2024)Vaccaro, Almaatouq, and Malone}]{vaccaro2024combinations}
Michelle Vaccaro, Abdullah Almaatouq, and Thomas Malone. 2024.
\newblock \href {https://doi.org/10.1038/s41562-024-02024-1} {{When combinations of humans and AI are useful: A systematic review and meta-analysis}}.
\newblock \emph{Nature Human Behaviour}, 8:2293--2303.

\bibitem[{Wadhwa et~al.(2024)Wadhwa, Zhao, Li, and Durrett}]{wadhwa2024refine}
Manya Wadhwa, Xinyu Zhao, Junyi~Jessy Li, and Greg Durrett. 2024.
\newblock \href {https://doi.org/10.18653/v1/2024.findings-emnlp.716} {{Learning to Refine with Fine-Grained Natural Language Feedback}}.
\newblock In \emph{Findings of the Association for Computational Linguistics: EMNLP 2024}, pages 12281--12308.

\bibitem[{Wang et~al.(2024)Wang, Wang, Liu, Chen, Yuan, Peng, and Ji}]{wang2024mint}
Xingyao Wang, Zihan Wang, Jiateng Liu, Yangyi Chen, Lifan Yuan, Hao Peng, and Heng Ji. 2024.
\newblock \href {https://proceedings.iclr.cc/paper_files/paper/2024/hash/8a0d3ae989a382ce6e50312bc35bf7e1-Abstract-Conference.html} {{MINT}: Evaluating {LLMs} in multi-turn interaction with tools and language feedback}.
\newblock In \emph{The Twelfth International Conference on Learning Representations}.

\bibitem[{Xiao et~al.(2024)Xiao, Treude, Hata, and Matsumoto}]{xiao2024devgpt}
Tao Xiao, Christoph Treude, Hideaki Hata, and Kenichi Matsumoto. 2024.
\newblock \href {https://doi.org/10.1145/3643991.3648400} {{DevGPT}: Studying developer-{ChatGPT} conversations}.
\newblock In \emph{Proceedings of the 21st International Conference on Mining Software Repositories}.

\bibitem[{Xie et~al.(2026)Xie, Qi, Yi, Yang, Whalen, Huang, Ding, Xie, Xie, and Wu}]{xie2026human}
Yueqi Xie, Tao Qi, Jingwei Yi, Xiyuan Yang, Ryan Whalen, Junming Huang, Qian Ding, Yu~Xie, Xing Xie, and Fangzhao Wu. 2026.
\newblock \href {https://doi.org/10.18653/v1/2026.acl-long.279} {Measuring human contribution in {AI}-assisted content generation}.
\newblock In \emph{Proceedings of the 64th Annual Meeting of the Association for Computational Linguistics (Volume 1: Long Papers)}, pages 6168--6190. Association for Computational Linguistics.

\bibitem[{Yang et~al.(2024)Yang, Yang, Zhang, Hui, Zheng, Yu et~al.}]{yang2024qwen25}
An~Yang, Baosong Yang, Beichen Zhang, Binyuan Hui, Bo~Zheng, Bowen Yu, et~al. 2024.
\newblock \href {https://arxiv.org/abs/2412.15115} {{Qwen2.5} technical report}.
\newblock \emph{Preprint}, arXiv:2412.15115.

\bibitem[{Yao et~al.(2025)Yao, Shinn, Razavi, and Narasimhan}]{yao2025taubench}
Shunyu Yao, Noah Shinn, Pedram Razavi, and Karthik Narasimhan. 2025.
\newblock \href {https://openreview.net/forum?id=roNSXZpUDN} {{{$\tau$}-bench: A Benchmark for Tool-Agent-User Interaction in Real-World Domains}}.
\newblock In \emph{The Thirteenth International Conference on Learning Representations}.

\bibitem[{Yao et~al.(2023)Yao, Zhao, Yu, Du, Shafran, Narasimhan, and Cao}]{yao2023react}
Shunyu Yao, Jeffrey Zhao, Dian Yu, Nan Du, Izhak Shafran, Karthik Narasimhan, and Yuan Cao. 2023.
\newblock \href {https://arxiv.org/abs/2210.03629} {{ReAct: Synergizing Reasoning and Acting in Language Models}}.
\newblock In \emph{The Eleventh International Conference on Learning Representations}.

\bibitem[{{Z.ai}(2026)}]{zai2026glm47flash}
{Z.ai}. 2026.
\newblock \href {https://huggingface.co/zai-org/GLM-4.7-Flash} {{GLM-4.7-Flash}: Official model card}.
\newblock Hugging Face.

\bibitem[{Zhao et~al.(2024)Zhao, Ren, Hessel, Cardie, Choi, and Deng}]{zhao2024wildchat}
Wenting Zhao, Xiang Ren, Jack Hessel, Claire Cardie, Yejin Choi, and Yuntian Deng. 2024.
\newblock \href {https://proceedings.iclr.cc/paper_files/paper/2024/hash/9421261e06f1a63a352b068f1ac90609-Abstract-Conference.html} {{WildChat}: 1m {ChatGPT} interaction logs in the wild}.
\newblock In \emph{The Twelfth International Conference on Learning Representations}.

\bibitem[{Zheng et~al.(2023)Zheng, Chiang, Sheng, Zhuang, Wu, Zhuang, Lin, Li, Li, Xing, Zhang, Gonzalez, and Stoica}]{zheng2023judge}
Lianmin Zheng, Wei-Lin Chiang, Ying Sheng, Siyuan Zhuang, Zhanghao Wu, Yonghao Zhuang, Zi~Lin, Zhuohan Li, Dacheng Li, Eric~P. Xing, Hao Zhang, Joseph~E. Gonzalez, and Ion Stoica. 2023.
\newblock \href {https://proceedings.neurips.cc/paper_files/paper/2023/hash/91f18a1287b398d378ef22505bf41832-Abstract-Datasets_and_Benchmarks.html} {{Judging LLM-as-a-Judge with MT-Bench and Chatbot Arena}}.
\newblock In \emph{Advances in Neural Information Processing Systems}, volume~36.

\bibitem[{Zhou et~al.(2024{\natexlab{a}})Zhou, Xu, Zhu, Zhou, Lo, Sridhar, Cheng, Ou, Bisk, Fried, Alon, and Neubig}]{zhou2024webarena}
Shuyan Zhou, Frank~F. Xu, Hao Zhu, Xuhui Zhou, Robert Lo, Abishek Sridhar, Xianyi Cheng, Tianyue Ou, Yonatan Bisk, Daniel Fried, Uri Alon, and Graham Neubig. 2024{\natexlab{a}}.
\newblock \href {https://openreview.net/forum?id=oKn9c6ytLx} {{WebArena: A Realistic Web Environment for Building Autonomous Agents}}.
\newblock In \emph{The Twelfth International Conference on Learning Representations}.

\bibitem[{Zhou et~al.(2024{\natexlab{b}})Zhou, Zhu, Mathur, Zhang, Yu, Qi, Morency, Bisk, Fried, Neubig, and Sap}]{zhou2024sotopia}
Xuhui Zhou, Hao Zhu, Leena Mathur, Ruohong Zhang, Haofei Yu, Zhengyang Qi, Louis-Philippe Morency, Yonatan Bisk, Daniel Fried, Graham Neubig, and Maarten Sap. 2024{\natexlab{b}}.
\newblock \href {https://arxiv.org/abs/2310.11667} {{SOTOPIA: Interactive Evaluation for Social Intelligence in Language Agents}}.
\newblock In \emph{The Twelfth International Conference on Learning Representations}.

\end{thebibliography}
